\documentclass[twocolumn]{aastex62}
\usepackage{natbib}
\usepackage{amsmath}
\usepackage{afterpage}

\begin{document}

\title{Probing the Influence of a Tachocline in Simulated M-Dwarf Dynamos}
\author{C.P. Bice}
\affiliation{JILA and Department of Astrophysical and Planetary Sciences, University of Colorado Boulder}
\author{J. Toomre}
\affiliation{JILA and Department of Astrophysical and Planetary Sciences, University of Colorado Boulder}

\correspondingauthor{Bice, Connor P.}
\email{connor.bice@colorado.edu}

\begin{abstract}
M-type stars are among the best candidates in searches for habitable Earth-like exoplanets, and yet many M-dwarfs exhibit extraordinary flaring which would bombard otherwise habitable planets with ionizing radiation. Observers have found that the fraction of M-stars demonstrating significant activity transitions from roughly 10\% for main-sequence stars more massive than 0.35 $M_\odot$, to nearly 90\% for less massive stars. The latter are typically rotating quite rapidly, suggesting differing spin-down histories. It is also below 0.35 $M_\odot$ that main-sequence stars become fully convective, and may no longer contain a tachocline. We turn here to the more massive M-stars to study the impact such a layer may have on their internal dynamics. Using the global MHD code Rayleigh, we compare the properties of convective dynamos generated within rapidly rotating 0.4 $M_\odot$ stars, with the computational domain either terminating at the base of the convection zone or permitting overshoot into the underlying stable region. We find that a tachocline is not necessary for the organization of strong toroidal wreaths of magnetism in these stars, though it can increase the coupling of mean field amplitudes to the stellar rotation rate. Additionally, we note that the presence of a tachocline tends to make magnetic cycles more regular than they would have otherwise been, and can permit alternative field configurations with much longer cycles. Finally, we find that the tachocline helps enhance the emergent fields and organize them into larger spatial scales, providing favorable conditions for more rapid spin-down via the stellar wind.
\end{abstract}

\keywords{convection, dynamo, MHD, stars: interiors, stars: low-mass, stars: magnetic field}

\section{Introduction}
M-dwarfs are quickly stepping into the forefront as some of the best candidates in modern searches for habitable, Earth-like exoplanets \citep{ballard}. This is due mainly to their small masses and luminosities, favoring close-in Goldilocks zones which translate to stronger and more frequent signals for many exoplanet detection schemes \citep{scalo}. The Goldilocks zone may not provide the whole picture for habitability, however, as many M-dwarfs exhibit extraordinary flaring events \citep{flares} which may bombard these exoplanets with ionizing radiation \citep{france}. 

These magnetic flares are reasonably interesting phenomenae in their own right, even disregarding their impacts on exoplanets. How is it that such dim, cool stars can produce flares whose luminosities exceed their brightest counterparts on the Sun by orders of magnitude? To begin to answer these questions, we turn to the convective dynamos operating in the interiors of these stars, where the roots of any flare must be formed. 

\subsection{Observational Motivations}
It has been clear for some time observationally that the magnetic activity a star is capable of generating is closely tied to its rotation rate \citep{rotact}. Under this rotation-activity relation, relatively slowly rotating stars follow a power-law whereby markers for magnetic activity increase with decreasing Rossby number, defined as the ratio of the rotational and convective timescales (faster rotation). Below a threshold of roughly R$_\mathrm{o}=0.1$, the magnetism seems to saturate. While rotation seems to be a crucial ingredient for yielding stellar magnetic fields with greater strengths, there is still room for substantial variation in the configuration of these fields with other stellar parameters.

Considering the activity of late-type M-stars using markers such as chromospheric emission of H$\alpha$, a sharp transition can be seen at roughly M3.5 ($0.35 M_\odot$) \citep{tachodiv}. Stars of spectral type earlier (more massive) than M3.5 are dominantly inactive, with only $10\%$ demonstrating significant markers for magnetism. Among main sequence stars later than M3.5, however, this fraction quickly grows to nearly $90\%$ of stars appearing to be magnetically active. Suggestively, stellar modeling tells us that it is later than M3.5 (below $0.35 M_\odot$) where main-squence stars become fully convective (FC). 

Among other things, becoming FC necessitates the loss of the transition region between the differentially-rotating convection zone (CZ) and the underlying  radiative zone (RZ) which may rotate as a solid body. Helioseismology tells us that within the Sun, this transition is a layer of substantial rotational shear and thus it has come to be called the tachocline \citep{spiegelz}. The stably-stratified shearing flows of the solar tachocline are often considered to be fundamental in organizing the Sun's magnetic fields for various dynamo theories (e.g., \citet{babcock}; \citet{ossenmft}; \citet{charbmft}). We seek to understand here how their presence or absence may be contributing to the striking divide observed between the distribution of magnetic activity among early and late type M-dwarfs.

\subsection{Magnetized Stellar Spin-Down}
The most probable explanation for this tachocline divide in magnetic behavior is not that the late M-dwarfs are just far more efficient at generating fields. Rather, this disparity is most likely the result of different evolutionary paths for FC and non-FC stars. Nearly all stars are likely born rapidly rotating and thus have the potential to be strongly magnetized, whereafter they shed angular momentum and gradually spin down. 

For relatively cool stars with near-surface CZs, the primary mechanism for angular momentum loss is through torques applied by a magnetized stellar wind \citep{parker}. The magnitudes of these torques are dependent on stellar parameters such as the mass, radius, rotation rate, and mass loss rate, as well as the strength and configuration of the emergent field from the star \citep{winds}. While stronger fields enable greater torques, just as important is to consider how far the stellar magnetic fields reach into the outflowing winds. In a multipole expansion, the radial decay of a magnetic field varies as $r^{-(l+1)}$, with $l$ the spherical harmonic degree, and thus only the low-order modes are capable of contributing a torque with a significant lever-arm.

Here we search for a potential source for differing evolutionary paths across the tachocline divide. If rapidly-rotating early M-dwarfs are able to spin-down more efficiently than their FC counterparts, possibly through greater field strengths or a field configuration favoring low-order modes, then their rotation-dependent magnetism would die out more quickly and leave behind a distribution of M-dwarf activity possibly resembling that which can be observed today.

\subsection{Features of Recent Dynamo Studies}
Global dynamo simulations attempting to recreate the solar CZ-tachocline dynamo system have yielded promising results. \citet{browning06} found that the tachocline could produce large-scale axisymmetric fields, which was historically difficult to achieve without one (e.g., \citealt{brun04}), while later work was able to achieve a cycling dynamo in the tachocline even at modest resolutions \citep{ghizaru10}. While the tachocline was once thought to be critical to the solar dynamo, simulations have shown that a solar-like CZ can, if rapidly rotating and sufficiently turbulent, sustain globally organized and periodically cycling \it wreaths \rm of magnetism even in the absence of a tachocline. These wreaths can be statistically steady \citep{brownsteady} or go through periodic cycles \citep{browncycle}, among a variety of other behavior stemming from an intricate and nonlinear parameter space. Similar results have been found for models of F-type stars containing tachoclines \citep{kyle13}, with slower rotation leading to magnetic cycles while faster rotation yielded steady fields. 

Comparing solar-like global simulations with and without the tachocline, \citet{guerrero16} found that in certain regions of parameter space the tachocline can dominate the dynamo action occurring in the CZ, yielding more much longer reversal time and better synchronicity between hemispheres. Models employing alternative treatments of the sub-gridscale (SGS) turbulence such as slope-limited diffusion have been able to stand out by recovering the enduring regularity and equatorward propagation of active latitudes that characterize the cycles of the real Sun \citep{kyle15}. Recently, such features have also been reproduced in a solar model with a more basic SGS treatment \citep{loren}. As technological advancements have brought smaller diffusivities into the reach of global dynamo simulations, we are beginning to see flux emergence as a general feature of these models (e.g., \citealt{nelson11}; \citealt{nelson13}; \citealt{nelson14}; \citealt{fanfang}), reaffirming the claim that the surface fields are intimately and directly linked to the internal dynamo.

In the realm of M-dwarfs, several global dynamo simulations of FC stars (e.g., \citealt{browning08}; \citealt{yadav15}) have demonstrated very strong magnetism reaching mean toroidal field strengths in excess of 10kG. The simulations with particularly strong fields appear to damp away nearly all of the differential rotation achieved by their hydrodynamic precursors. In Yadav, broad dipolar caps of magnetism powerful enough to partly suppress convection and create polar dark spots were produced. Though vastly different from M-dwarfs in overall stellar properties, the convection taking place in the cores of A-type stars (\citealt{brun05}, \citealt{feather09}) and massive B-stars \citep{kyle16} occurs in a similar geometry to that of the FC late M-dwarfs. Similarly to the FC models, these studies found that the majority of the magnetic energy was non-axisymmetric, and the differential rotation was nearly completely quenched. 

\section{Framing the Problem}
We turn now to study early M-dwarfs featuring deep CZs and underlying RZs, considering dynamics attained within simulations of full spherical shells of electrically conducting fluid. We employ the open-source 3D MHD code Rayleigh \citep{rayleigh} to evolve the anelastic compressible equations in rotating spherical shells. Rayleigh is a pseudospectral code, employing both a physical grid and a basis of spherical harmonics and Chebyshev polynomials. Time stepping is achieved with a hybrid implicit-explicit approach, where the linear and nonlinear terms are advanced in spectral space via 2nd-order Crank-Nicolson and in physical space by a 2nd-order Adams-Bashforth method, respectively.
\subsection{The Anelastic Equations}
The anelastic equations are a fully nonlinear form of the fluid equations from which sound waves have been filtered out. This provides an appropriate framework for exploring subsonic convection within stellar interiors, where fast-moving p-modes would otherwise throttle the maximum timestep. The thermodynamic variables are linearized against a one-dimensional, time independent background state involving density, pressure, temperature, and entropy ($\bar{\rho},\,\bar{P},\,\bar{T},$ and $\bar{S}$, respectively), with deviations from the background written without overbars. Due to the resolutions accessible to modern computing, the viscosity $\nu$, conductivity $\kappa$, and resistivity $\eta$ we employ are not the molecular values, but rather eddy diffusivities. These values are inflated by many orders of magnitude as a parameterization of the turbulent mixing occurring at sub-grid scales. The detailed forms of the anelastic equations involving the velocity vector $\mathbf{v}$ and the magnetic field vector $\mathbf{B}$ solved in Rayleigh are as follows.

\begin{equation}
\begin{split}
\mathrm{Mome}&\mathrm{ntum:}\;\;\bar{\rho}(\frac{D\mathbf{v}}{Dt}+2\Omega_0\hat{z}\times\mathbf{v})=\\&-\bar{\rho}\nabla\frac{P}{\bar{\rho}}+ \frac{\bar{\rho}g}{c_p}S+\nabla\cdot\mathcal{D}+\frac{1}{4\pi}(\nabla\times\mathbf{B})\times\mathbf{B}\;,
\end{split}
\label{eqn:momentum}
\end{equation}
\begin{equation}
\begin{split}
\mathrm{Energy:}\;\;\bar{\rho}\bar{T}\frac{DS}{Dt}&=\nabla\cdot[\kappa\bar{\rho}\bar{T}\nabla S]+\\&Q+\Phi+\frac{\eta}{4\pi}[\nabla\times\mathbf{B}]^2,
\end{split}
\label{eqn:thermal}
\end{equation}
\begin{equation}
\mathrm{Induction:}\;\;\frac{\partial\mathbf{B}}{\partial t}=\nabla\times(\mathbf{v}\times\mathbf{B}-\eta\nabla\times\mathbf{B})
\label{eqn:induction}
\end{equation}
\begin{equation}
\mathrm{Continuity:}\;\;\nabla\cdot(\bar{\rho}\mathbf{v})=0\;,
\label{eqn:continuity}
\end{equation}
\begin{equation}
\mathrm{Solenoidal:}\;\;\nabla\cdot\mathbf{B}=0\;.
\end{equation}
Here, $Q$ is the volumetric heating function, $\mathcal{D}$ is the viscous stress tensor, and $\Phi$ represents the viscous heating, which are defined as

\begin{equation}
\mathcal{D}_{ij}=2\bar{\rho}\nu[e_{ij}-\frac{1}{3}(\nabla\cdot\mathbf{v})]\;.
\end{equation} 
\begin{equation}
\Phi=2\bar{\rho}\nu[e_{ij}e_{ij}-\frac{1}{3}(\nabla\cdot\mathbf{v})^2]\;,
\end{equation} 
with $e_{ij}$ as the strain rate tensor. Closure is achieved with a linearized equation of state, 

\begin{equation}
\frac{P}{\bar{P}}=\frac{\rho}{\bar{\rho}}+\frac{T}{\bar{T}}\;.
\end{equation}

\subsection{Modeling an M-Dwarf}

\begin{figure}
	\centering
	\includegraphics[width=1.0\linewidth]{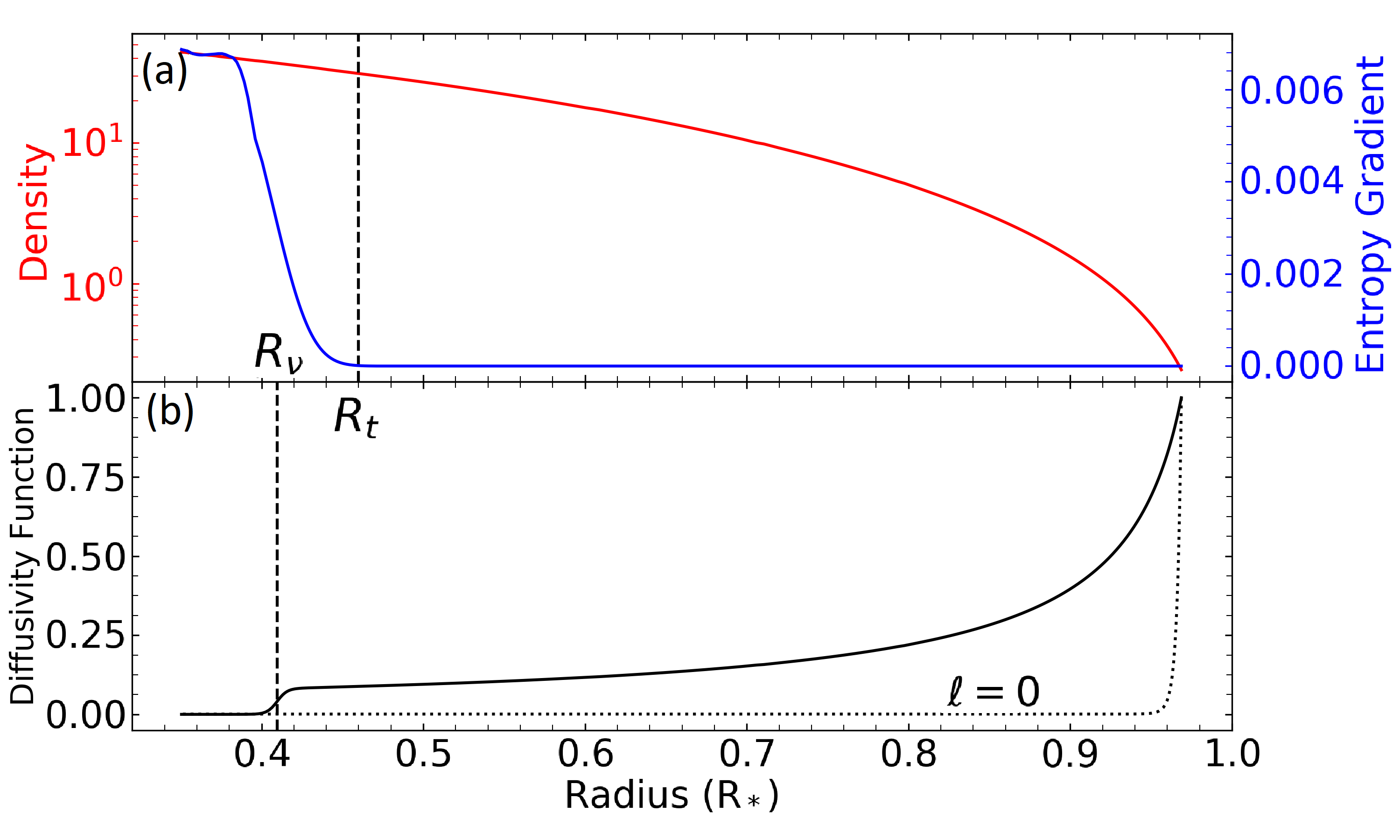}
	\caption{(a) The density stratification (red) and background entropy gradient (blue) employed by the simulations in proportional radius. A smoothing function applied to the entropy profile results in a somewhat more gradual transition to convective stability here than indicated by the stellar model. $R_t$ is marked with a vertical dashed line, while $R_i$ and $R_o$ are at the endpoints of the profiles. (b) The shapes of the diffusion profiles used in the simulation. Most coefficients are drawn from the solid curve, however the $l=0$ component of the thermal conductivity uses the dotted. The radii of the diffusive transition $R_\nu$ is marked with a vertical dashed line.}
	\label{fig:reference}
\end{figure}

The calculations were performed within a radial hydrodynamic background state derived using the stellar evolution community code MESA \citep{MESA}. We consider a ZAMS star of 0.4 M$_\odot$ with solar metalicity, a luminosity of $9.478\times10^{31}$ erg s$^{-1}$ ($0.025 L_\odot$), and rotating at 0.5, 1, 2, and 4 times the solar rate, $\Omega_\odot=414\,\mathrm{nHz}$, corresponding to rotation periods of 55.6, 27.8, 13.9, and 7.0 days. The chosen rotation rates yield Rossby numbers (defined as $\mathrm{R_o}=v_{rms}/(\Omega_*L)$, with $L$ the depth of the CZ) which would place these models in the saturated (strong field) regime of the rotation-activity relation. In the outermost layers of stars, the anelastic equations begin to break down as flows approach the sound speed and non-diffusive radiative transfer becomes important. As a result, we must restrict our computational domain to exclude this region. A plot of the density stratification and entropy gradient is presented in Figure \ref{fig:reference}a. All simulations had CZs extending from $R_t= 0.42R_*$ to $R_o=0.97R_*$, where $R_*=2.588\times 10^{10}$ cm, spanning $N_\rho=5$ density scale heights. The tachocline models contained an additional radial domain spanning the tachocline region and underlying stable layer, $R_i=0.35R_*$ to $R_t$. \textbf{This construction ensures that any differences which emerge between the two classes of models can be attributed dominantly to the presence or absence of a tachocline, as their CZs are nearly identical.} A list of all models and their parameters is given in Table \ref{tab:flows}. The non-dimensional numbers presented there are the Rayleigh number, defined as $\mathrm{R_a}=(gQL^5)/(c_p\nu\kappa^2)$, the Taylor number $\mathrm{T_a}=(4\Omega^2L^4)/\nu^2$, and the Reynolds numbers $\mathrm{R_e}=(u_{rms}L)/\nu$ and $\mathrm{R_e}'=((u-\langle u\rangle_\phi)_{rms}L)/\nu$. 

As used in many previous studies (e.g., \citealt{brun04}; \citealt{brownsteady}; \citealt{kyle13}; \citealt{fanfang}), we employ viscosity $\nu$ profiles for our non-tachocline models proportional to $\bar{\rho}^{-\frac{1}{2}}$ where the viscosity at the top of the domain is chosen to be $\nu_0=6.65\times 10^{11} $ cm$^2 $s$^{-1}$. The thermal conductivity $\kappa$ and resistivity $\eta$ have the same profiles and are chosen to yield a Prandtl number P$_\mathrm{r}=\nu/\kappa=0.25$ and magnetic Prandtl number P$_\mathrm{rm}=\nu/\eta=4$ throughout the domain, except in case D2a where we set $\mathrm{P_{rm}}=1$. A scaling study for non-rotating spherical anelastic convection performed by \citet{rayleigh} suggests that the kinetic energy density of the flows saturates above a flux Rayleigh number roughly 100 times the critical value. Accordingly, we have chosen our diffusivities to place all of our models firmly within this regime (see Table \ref{tab:flows}). 

Although substantive attempts have been made to explain the striking thinness of the solar tachocline and its apparent longevity  (e.g., \citealt{spiegelz}; \citealt{elliot97}; \citealt{goughm}; see summary in \citealt{LRtacho}), there is still no clear consensus as to the mechanism by which it is maintained. Accordingly, our diffusion profiles for models containing stable layers have identical structure in the CZ to their purely convective counterparts, but sharply reduce their amplitudes in the tachocline. This reflects the theoretical expectation of reduced turbulent mixing in the stable region, and serves to increase the viscous time scale there, delaying the eventual unravelling of the shear layer as the differential rotation of the CZ imprints downward. The functional form of this diffusion profile is

\begin{equation}
\nu = \nu_t+\frac{\nu_0(\frac{\rho}{\rho_0})^{-0.5}}{1+\exp{(c(R_\nu-r)/(R_o-r_i))}}\;.
\end{equation}

Here we choose tachocline diffusivities $\nu_t=10^{-4}\nu_0$ and a transition steepness $c=200$, yielding a viscous timecale in the tachocline of approximately $t_{visc} = 1100$ years. The functional forms of $\kappa$ and $\eta$ are the same. In all simulations, the mean $(l=0)$ entropy field sees a separate conductivity $\kappa_m$, defined as 

\begin{equation}
\kappa_m = \kappa_t+(\kappa_0-\kappa_t)(\frac{\rho}{\rho_0})^{-p};.
\end{equation}

We choose $p=8$, which causes $\kappa_m$ to drop to its floor value a short distance below the top boundary. This serves to discourage thermal conduction as a means of energy transport in the bulk of the CZ, and consequently forces the convective motions to carry nearly the full luminosity of the star. Non-dimensional forms of both diffusivity profiles are presented in Figure \ref{fig:reference}b.
 
\subsubsection{Boundary Conditions} 
\begin{table*}[]
\centering
\begin{tabular*}{\textwidth}{l @{\extracolsep{\fill}} cccccccccc}
\hline \hline
Case & $N_r\times N_\theta\times N_\phi$ & $\Omega / \Omega_\odot$ & $\nu_0$ & $\mathrm{P_{rm}}$ & $\mathrm{R_a} / \mathrm{R_{a,crit}}$ & $\mathrm{T_a}/10^8$ & $\mathrm{R_e}$ & $\mathrm{R_e}'$ & $\mathrm{R_o}/10^{-1}$ & $\Delta\Omega$    \\ \hline
H05t & $(48+192)\times384\times768$ & 0.5 & 6.65 & - & $2.28\times10^4$ & 0.23 & 237 & 124 & 10.8 & 54.7\% \\
H1t & $(48+192)\times384\times768$ & 1 & 6.65 & - & $4.84\times10^3$ & 0.90 & 271 & 95.9 & 6.18 & 37.6\% \\
H2 & $192\times384\times768$ & 2 & 6.65 & - & $1.25\times10^3$ & 4.19 & 263 & 85.8 & 2.57 & 16.0\%\\
H2t & $(48+192)\times384\times768$ & 2 & 6.65 & - & $1.03\times10^3$ & 3.60 & 284 & 80.4 & 2.99 & 18.6\%\\
H4t & $(48+192)\times384\times768$ & 4 & 6.65 & - & $2.19\times10^3$ & 14.4 & 230 & 63.4 & 1.31 & 7.6\%\\ \hline
D1 & $192\times384\times768$ & 1 & 6.65 & 4 & $5.85\times10^3$ & 1.05 & 132 & 109 & 2.58 & 9.1\%\\
D1t & $(48+192)\times384\times768$ & 1 & 6.65 & 4 & $4.85\times10^3$ & 0.901 & 145 & 96.2 & 3.29 & 3.7\%\\
D2 & $192\times384\times768$ & 2 & 6.65 & 4 & $1.25\times10^3$ & 4.19 & 99.9 & 92.6 & 0.977 & 3.2\%\\
D2t & $(48+192)\times384\times768$ & 2 & 6.65 & 4 & $1.03\times10^3$ & 3.60 & 127 & 89.6 & 1.34 & 3.8\%\\
D2a & $192\times384\times768$ & 2 & 6.65 & 1 & $1.25\times10^3$ & 4.19 & 122 & 96.1 & 1.19 & 4.7\%\\
D2ta & $(48+192)\times384\times768$ & 2 & 6.65 & 1 & $1.03\times10^3$ & 3.60 & 162 & 96.6 & 1.73 & 4.6\%\\
D4 & $192\times384\times768$ & 4 & 6.65 & 4 & $2.66\times10^2$ & 16.7 & 86.2 & 78.1 & 0.421 & 1.7\%\\
D4t & $(48+192)\times384\times768$ & 4 & 6.65 & 4 & $2.19\times10^2$ & 14.4 & 128 & 67.7 & 0.730 & 1.6\%\\ \hline \hline
\end{tabular*}
    \caption{Model parameters, along with time- and shell-averaged flow characteristics measured near mid-depth, $r=0.70R_*$. In our notation, models are either ``H" hydrodynamic or ``D" dynamo, followed by the frame rotation rate $\Omega_0$ in multiples of $\Omega_\odot$. Models whose name contains ``t" incorporate a tachocline in their computational domain, whereas those without are purely CZ; the suffix ``a" denotes a deviation from the standard diffusion parameters. N$_\theta=384$ corresponds to a spectral resolution of $\ell_\mathrm{max}=255$. Viscosities $\nu_0$ are reported at the top of the domain in units of $10^{11}$ cm$^2$s$^{-1}$. Rayleigh numbers are of the form $\mathrm{R_a}=(gQL^5)/(c_p\nu\kappa^2)$, which is appropriate for convection being driven by a distributed internal heating function $Q(r)$. Critical Rayleigh numbers are determined empirically and reflect the influence of rotation but not magnetism. The Taylor number is defined as $\mathrm{T_a}=(4\Omega^2L^4)/\nu^2$. The Reynolds number is $\mathrm{R_e}=(u_{rms}L)/\nu$ and the fluctuating Reynolds number is $\mathrm{R_e}'=((u-\langle u\rangle_\phi)_{rms}L)/\nu$. The Rossby number is given by $\mathrm{R_o}=v_{rms}/(\Omega_*L)$. Finally, the amplitude of the differential rotation is reported as a percentage of the frame rotation rate $\Delta\Omega = (\Omega_{eq}-\Omega_{75})/\Omega_*$.}
    \label{tab:flows}
\end{table*}
The upper and lower boundary conditions are impenetrable and stress free, 
\begin{equation}
v_r|_{\mathrm{bc}}=\frac{d}{dr}(v_\theta/r)|_{\mathrm{bc}}=\frac{d}{dr}(v_\phi/r)|_{\mathrm{bc}}=0\;.
\end{equation}

The lower boundary is thermally insulating, and the top boundary extracts the star's luminosity through a fixed conductive gradient, with
\begin{equation}
\frac{dS}{dr}|_{\mathrm{bot}}=0,\; \frac{dS}{dr}|_{\mathrm{top}}=\mathrm{const}\;.
\end{equation}

With no conductive input, energy balance is instead maintained through the volumetric heating function $Q$ which is adapted from the $\epsilon_{nuc}$ and $\nabla\cdot\mathcal{F}_{rad}$ reported by MESA. Finally, the magnetic field matches onto an external potential field at both boundaries, as
\begin{equation}
B=\nabla\Phi,\; \nabla^2\Phi|_{R_i,R_o}=0\;.
\end{equation}

Physically, it may be more appropriate to consider the bottom boundary to be a perfect conductor due to the eddy resistivities becoming very small, and indeed simulations using this condition such as \citet{brownsteady} have found that such a choice can enable stronger and more stable fields. We choose the potential field condition due to concerns that the most common implementations of the perfect conductor boundary may be overconstraining the fields.

The hydrodynamical models were evolved first, in the absence of magnetism, to study their general properties. \textbf{To ensure numerical convergence, these models are initiated with roughly double the resolution reported in Table \ref{tab:flows}, and only shifted to their current grids after passing through the transient of convective onset.} After a statistical steady state was achieved, magnetism was introduced as white-noise perturbations and allowed to self-consistently reshape the flows while growing to its mature amplitudes. Where equivalent hydrodynamical models were unavailable (as in cases D1 and D4), we instead branched these cases from a solution with identical geometry and diffusion but differing rotation rate. We note that similar parameter spaces in the solar regime are fraught with regions of multiple stability, and thus there is the possibility that a different path to maturity for these models may result in different flow or field configurations. \textbf{As a final check on numerical convergence, we have compared power spectra for the velocity and magnetic fields realized in our reported models to those of identical cases with roughly 50\% more resolution in each dimension, finding a tight agreement.}

\section{Flows Achieved}
We begin by examining the patterns of convection realized in our five hydrodynamic and eight MHD simulations, exploring the direct influence of a tachocline on our flows as well as their responses to the magnetic feedback at different rotation rates. Unless otherwise noted, time averages of the hydrodynamic properties of the models are calculated over all time following that model's maturation, be that rotational or magnetic. 

\subsection{Patterns of Hydrodynamic Convection}

\begin{figure*}[p]
	\centering
	\includegraphics[width=0.97\linewidth]{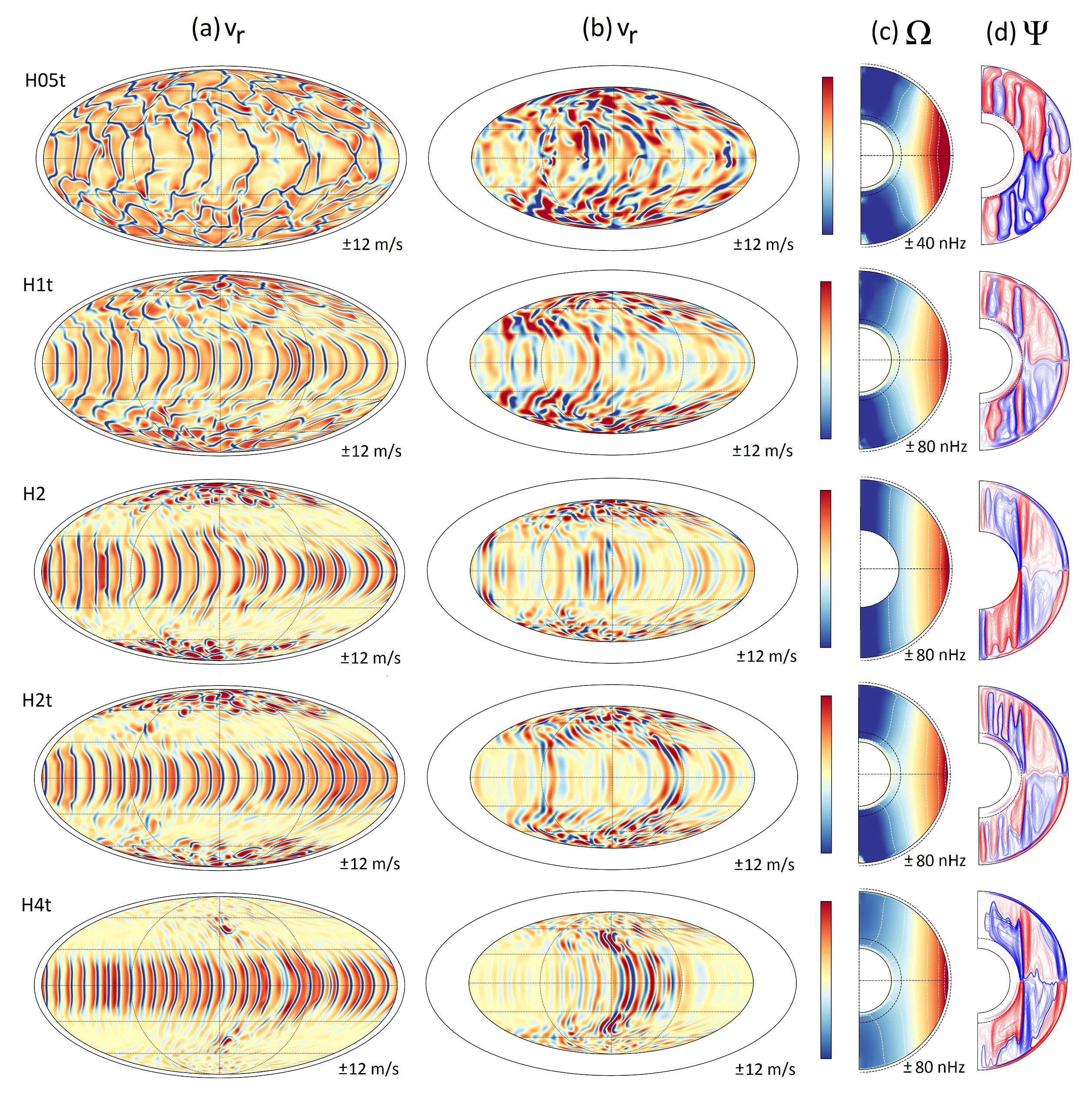}
	\caption{(a) Snapshots of radial velocity $v_r$ near-surface ($r=0.96R_*$) and (b) near mid-depth ($r=0.77R_*$), showing the full sphere in Mollweide projection (poles top and bottom, equator along center line) for each of the five hydrodynamic models. Colors vary from red (positive) to blue (negative) within the ranges at the bottom of each panel, here and in successive figures. A range of flow morphologies is evident in comparisons between cases with differing rotation rates. The outer surface of the star is marked with a black ring. (c) Time- and azimuthally-averaged differential rotation $\langle\Omega\rangle_{\phi,t}$ for each of the hydrodynamic cases, projected into the meridional plane. Dashed lines indicate the outer surface of the star and the transition to convective stability at the base of the CZ. (d) Time- and azimuthally-averaged mass flux $\Psi$ for each hydrodynamic case, showing the meridional circulations. Red (blue) lines denote a clockwise (counter-clockwise) circulation cell, while the density of lines encodes the total mass flux associated with a cell (high line density thus representing larger-scale and/or more vigorous circulations).}
	\label{fig:flows}
\end{figure*}

The variation of convective patterns across our hydrodynamic models is illustrated in columns (a) and (b) of Figure \ref{fig:flows}, showing radial velocity in Mollweide projection at two depths. All the flows are richly time-dependent, and here we present only representative snapshots. While there are many morphological distinctions to be drawn between our models, they are unified by an asymmetry between their broad, slow upflows, and narrow, fast downflows. This is a feature that can be found for nearly any compressible convection occurring in a stratified environment. Upflows carry material from regions that are hotter and denser than their new surroundings, and so they naturally tend to expand as they rise; the inverse is true for downflows. 

As the rotation rate increases, the nearly isotropic convection of case H05t gives way to a clear separation between rolling, axially-aligned structures known as Busse columns near the equator and more isotropic plumes near the poles. As the former extend away from the equator, the latitudinal shear of the differential rotation causes their outer edges to lag behind, giving them their characteristic banana-like shapes. This effect emerges as a consequence of the Taylor-Proudman theorem, which states that when rotational constraint is strong, or equivalently, when the Rossby number is small, flows tend to be uniform along columns parallel to the axis of rotation. Near-surface spherical cuts (Figure \ref{fig:flows}a) then allow us to see these columns along their edges near the equator, while we observe their tops at high-latitude. Cuts near mid-depth (Figure \ref{fig:flows}b) demonstrate the connectivity of the convective cells, which may span the full radial domain for the strongest flows. Considering a cylinder tangent to the stable region and aligned with the rotation axis, which meets the outer boundary of our domain at approximately $\pm 60^\circ$, there is a clear and substantive difference between flows at high- and low-latitudes under rotational constraint. Flows outside the tangent cylinder, near the equator, may meld with their counterparts from the opposite hemisphere. Due to the presence of the stable region, this is impossible for flows within the tangent cylinder. Consequently, the tangent cylinder is often an important boundary for the formation of structures in rotationally constrained systems.

The horizontal width of the Busse columns, and thus the number that may fit within our simulation, has been shown through linear analysis of the rotating Boussinesq equations \citep{dormy04} and numerical computation on the linearized anelastic equations (\citealt{gilmandglatz}; \citealt{jones09}) to depend on the Taylor number, $\mathrm{T_a}$. They find the azimuthal wavenumber of the most unstable mode to be $m=m_c\mathrm{T_a}^{1/6}$, where $m_c$ is the critical wavenumber, or $m\propto\Omega^{1/3}\nu^{-1/3}$. A list of $\mathrm{T_a}$ values and other flow characteristics for each case is presented in Table \ref{tab:flows}. Centering on the $m=20$ of case H1t, this suggests we should expect azimuthal wavenumbers of 16, 25, and 32 for cases H05t, H2t, and H4t, respectively. In solid agreement, the measured values are 14, 25, and 32.

\begin{figure}
\centering
\includegraphics[width=0.95\linewidth]{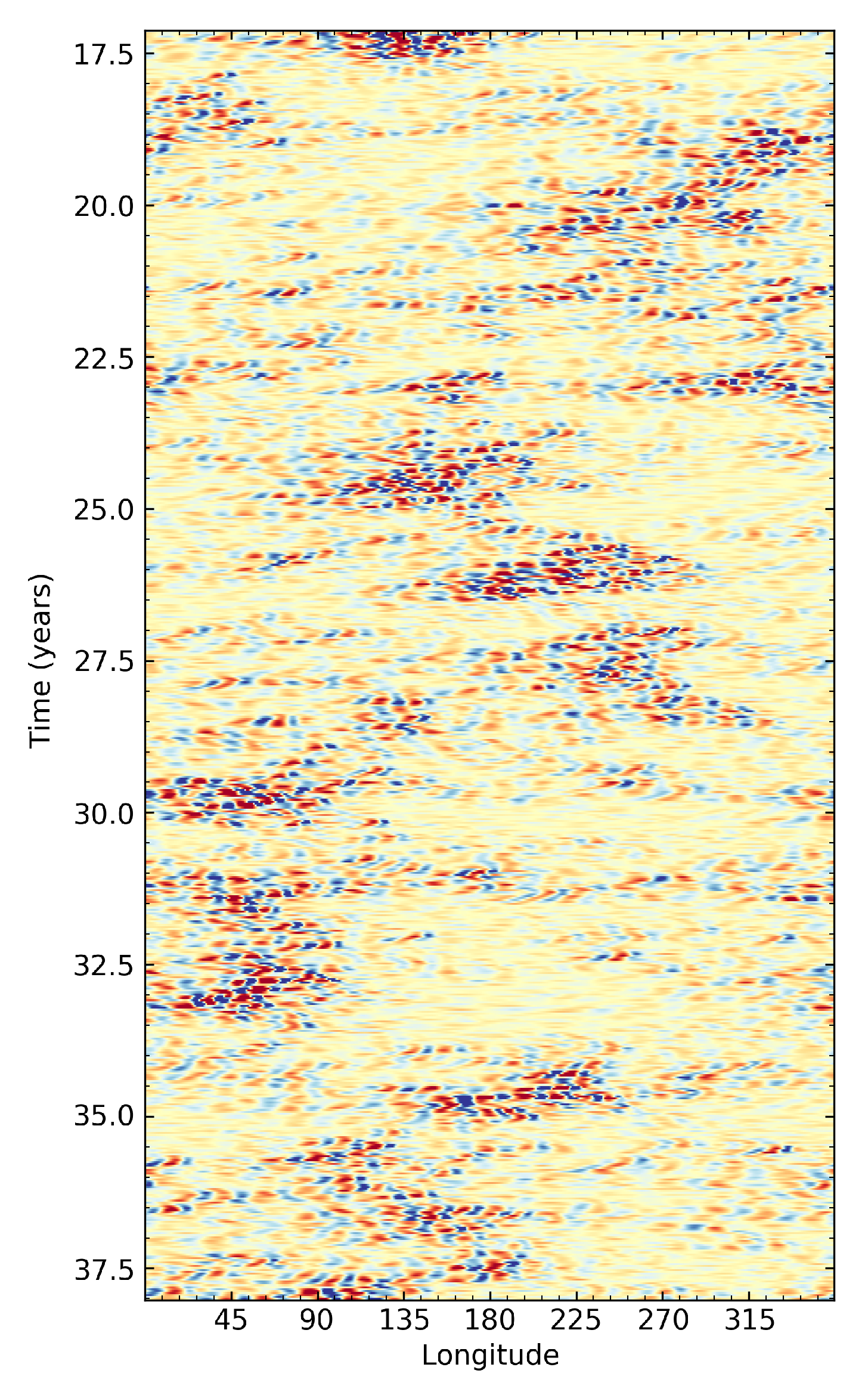}
\caption{Time-longitude diagram of $v_r$ in case H2t. Sampled along the equator near mid-depth and a tracking rate of 26.3 nHz prograde relative to the co-rotating frame. Intermittent patches of enhanced convection are evident and may propagate with rates greater or less than that of the chosen frame. Color coding is the same as in Figure \ref{fig:flows}.}
\label{fig:patchy}
\end{figure}

The influence of rotational constraint on convection is brought again to the forefront by the presence of \it localized nests \rm of stronger convection near the equator evident in the faster-rotating cases H2, H2t, and H4t. A time-longitude diagram of the mid-depth radial velocities of case H2t, shown in Figure \ref{fig:patchy}, reveals their intermittent structure. Spatially modulated convection has been observed in past simulations of the geodynamo \citep{groteandbusse} and the solar interior \citep{brown08}. Similarly to the results of Brown et al., the convective plumes associated with our active nests feature significantly greater radial extent than those in the outlying areas. The nests are coherent structures, which propagate prograde with angular velocities intermediate between the frame rotation and the local differential rotation, ranging from roughly 15 to 35 nHz for case H2t (1.8\% to 4.2\% of $\Omega_*$). When two nests are simultaneously present, differing angular velocities may bring them together, causing them to merge into a single nest of greater amplitude. It is rare for both nests to survive such an encounter and pass through each other, casting doubt on the interpretation that they may be solitonic in nature.

\begin{figure}
	\centering
	\includegraphics[width=1.0\linewidth]{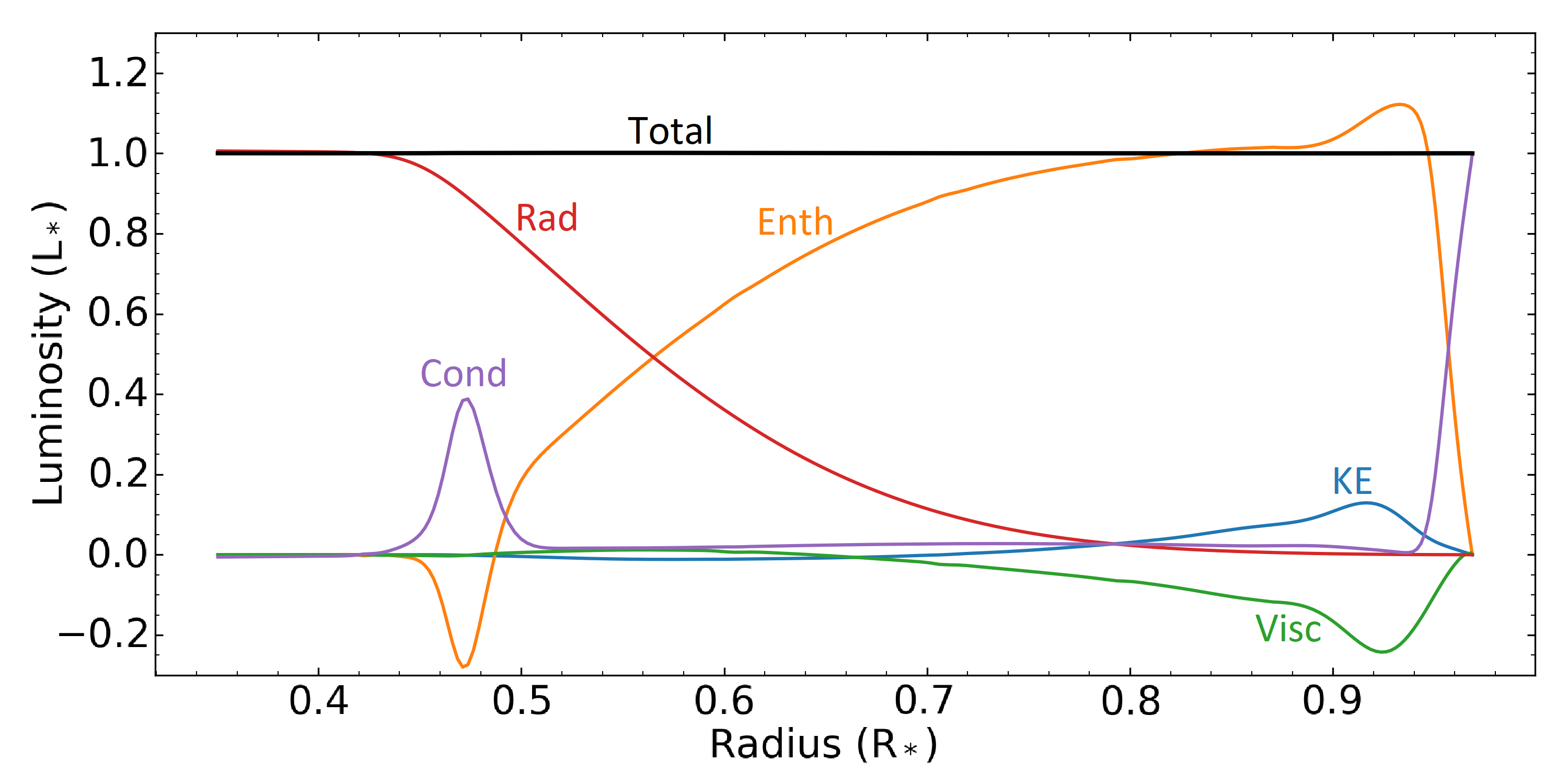}
	\caption{Energy flux balance in case H2t. An overluminous enthalpy flux near the surface is balanced by a negative viscous flux associated with the fast and narrow downflows. The enthalpy goes negative near the base of the CZ as convection overshoots into a stably stratified region, and is compensated by an enhanced thermal conduction there.}
	\label{fig:eflux}
\end{figure}

Considering the role of convection as the primary mode of energy transport through the CZ, such a modulation of its vigor may lead to asphericity in surface temperatures. Due to the thermal boundary conditions we impose, the amplitude of perturbations to the surface temperature in our models are limited. Despite this, however, we observe the formation of travelling hot spots above our nests, reaching a temperature enhancement of roughly 1 K relative to the equatorial mean in case H4t. Should these convective patches prove to be robust features of fast rotation in the parameter spaces of real stars, they may have interesting observational consequences. Similar to sunspots, a large-scale hotspot located along the equator could lead to intermittent variations in photometric or Doppler measurements of the stellar surface with cyclic periods on the order of the rotation period.

\subsection{Hydrodynamic Mean Structures}
The differential rotation $\langle\Omega-\Omega_*\rangle_{\phi,t}$ achieved in each hydrodynamic case is presented in column (c) of Figure \ref{fig:flows}. All models yield solar-like profiles, with fast equators and slow poles. Defining a rotational contrast between the equator and a latitude of $75^\circ$ at the surface, $\Delta\Omega=(\Omega_\mathrm{eq}-\Omega_{75})/\Omega_*$, we find contrasts of 54.7, 37.6, 16.0, 18.6, and 7.6\% for models H05t, H1t, H2, H2t, and H4t, respectively, as reported in Table \ref{tab:flows}. Considering the successive doublings of the frame rotation rate, this suggests that the amplitude of the differential rotation established is largely the same across each of our hydrodynamic models. In the presence of a tachocline, we find that the high-latitude rotation contours break from their rotational constraint and become more radially aligned, a familiar feature of the solar differential rotation. The radial shear of the tachocline persists in all hydrodynamic models at the times shown.

The meridional circulations achieved in each of our hydrodynamic cases are shown in Figure \ref{fig:flows}(d). The cases with faster rotation each developed thin, poleward near-surface circulations, overlying one large cell in each hemisphere outside the tangent cylinder. Within the tangent cylinder, axially-aligned cells of alternating circulation direction extend through the full depth of the convection zone, their number increasing with faster rotation. In case H05t, the near surface cells do not extend to high latitudes, nor do the circulations respect the tangent cylinder. 


\subsection{Energy and Momentum Transport}

\begin{figure}
	\centering
	\includegraphics[width=1\linewidth]{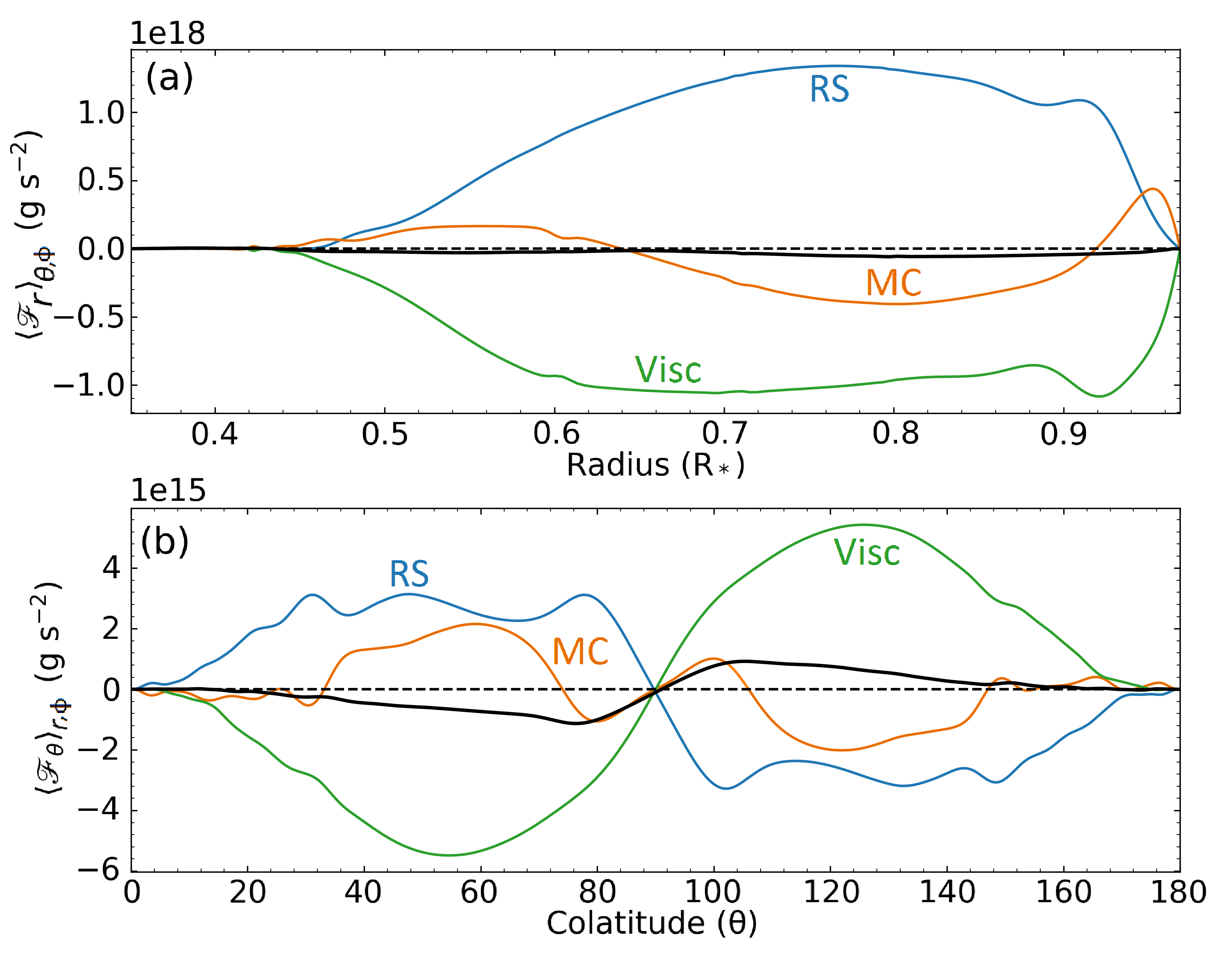}
	\caption{(a) Radial balance of spherically-averaged angular momentum flux for case H2t, which is dominated by an opposition between the Reynolds and viscous stresses. (b) Latitudinal balance of cone-averaged angular momentum flux for case H2t. Here the Reynolds stress and meridional circulation are co-aligned at mid-latitudes, and are again opposed by the viscous stress.}
	\label{fig:amflux}
\end{figure}

The flows achieved in our simulations transport energy and angular momentum throughout the computational domain. Due to their inherent complexity, these flows will not in general be well predicted by the 1-D mixing length arguments that were used to initialize our background state. The system finds a new self-consistent balance, whose mean state may differ from the initial conditions. Shown in Figure \ref{fig:eflux} for case H2t, this balance is maintained by five different energy fluxes associated with the kinetic energy, enthalpy, viscosity, conduction, and radiation, respectively. They are defined as

\begin{figure*}
	\centering
	\includegraphics[width=0.97\linewidth]{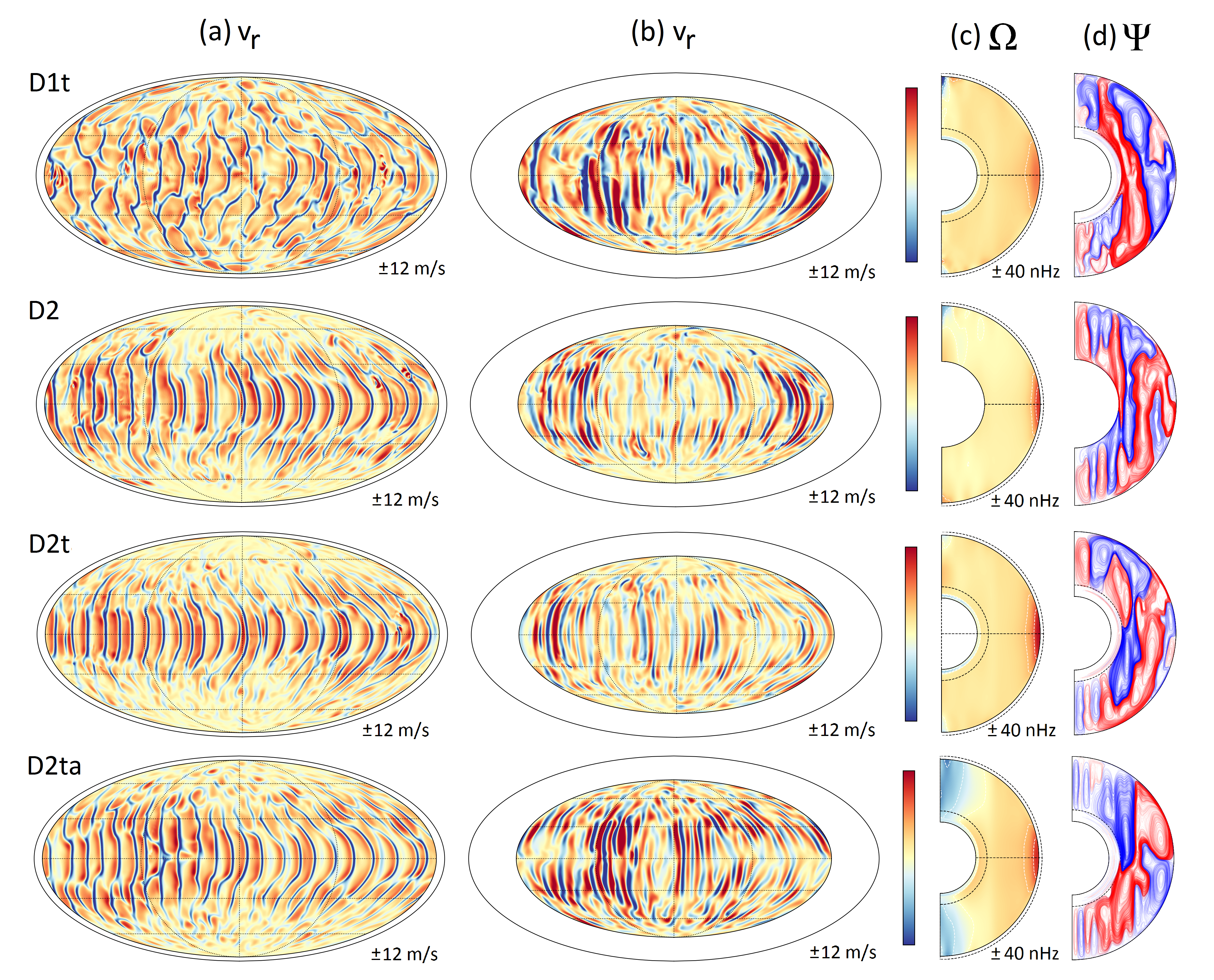}
	\caption{(a) Snapshots of radial velocity $v_r$ near surface ($r=0.96R_*$) and (b) near mid-depth ($r=0.77R_*$) for a selection of dynamo cases. The structure of the convection has been dramatically altered at mid to high latitudes through dynamo action. Patchy convection is evident near mid-depth in each case. (c) Differential rotation as shown in $\langle \Omega \rangle_{\phi,t}$ in a selection of dynamo cases, averaged in longitude and time. $\langle\Omega\rangle_{\phi,t}$ is strongly quenched by the dynamo action, except in cases D2a (not shown) and D2ta. (d) Meridional circulation for a selection of dynamo cases, averaged in longitude and time. Circulation cells are significantly disrupted by the influence of magnetism, often breaking their symmetry.}
	\label{fig:mhdconv}
\end{figure*}

\begin{equation}
\mathcal{F}_\mathrm{KE}=\frac{1}{2}\bar{\rho}v^2v_r\;,
\label{eqn:keflux}
\end{equation}
\begin{equation}
\mathcal{F}_\mathrm{Enth}=(\bar{\rho}\bar{T} S +\bar{P})v_r\;,
\end{equation}
\begin{equation}
\mathcal{F}_\mathrm{Visc}=-[\mathbf{v}\cdot\mathcal{D}]_r\;,
\end{equation}
\begin{equation}
\mathcal{F}_\mathrm{Cond}=-\kappa\bar{\rho}\bar{T}\frac{dS}{dr}\;.
\end{equation}

The radiative energy flux is not treated directly within the simulation. Instead, the time-invariant heating function $Q$ is designed to recreate the radiation field prescribed by MESA,
\begin{equation}
\mathcal{F}_\mathrm{Rad}=\frac{1}{4\pi r^2}[L_*-\int_{r_i}^r 4\pi r^2 \bar{\rho T}Qdr]\;.
\end{equation}

Radial $\mathcal{F}_\mathrm{Enth}$ becomes negative for flows in regions that are stable to convection, and so it is clear from Figure \ref{fig:eflux} that the base of the CZ in our tachocline containing simulations lies somewhere between $0.44R_*$ and $0.49R_*$, rather than at the $0.42R_*$ prescribed by our background state, and we adopt $R_{t}=0.46R_*$ as the canonical depth of the CZ for these models. The position of the top of the overshooting layer varies latitudinally, with the onset of convective stability lying roughly 2\% of the stellar radius deeper at the equator than at the poles. The bottom boundary of the overshoot region is nearly perfectly spherical. This self-adjustment is not possible for our CZ-only models, which remain convectively unstable throughout the entire domain, and so our comparisons between these two sets of cases must bear in mind a slight disparity in the depths of their CZs. Because the radiative flux is not allowed to evolve in time, a positive conductive flux emerges at the base of our CZ to balance the negative enthalpy flux there. 


The radial and latitudinal torque balance for case H2t is presented in Figure \ref{fig:amflux}, the various terms of which are defined for the Reynolds stress, meridional circulation, and viscous torque as

\begin{equation}
\langle \mathbf{\mathcal{F}_\mathrm{RS}}\rangle_\phi = \bar{\rho}r\sin\theta\langle \mathbf{v_m}'v_\phi'\rangle_\phi \;,
\end{equation}
\begin{equation}
\langle \mathbf{\mathcal{F}_\mathrm{MC}}\rangle_\phi = \bar{\rho}r\sin\theta\langle \mathbf{v_m}\rangle_\phi (\langle v_\phi\rangle_\phi +\Omega_0r\sin\theta)\;,
\end{equation}
\begin{equation}
\langle \mathbf{\mathcal{F}_\nu}\rangle_\phi = \bar{\rho}\nu[2\langle v_\phi\rangle_\phi \nabla(r\sin\theta)-\nabla(r\sin\theta \langle v_\phi\rangle_\phi )]\;,
\end{equation}
where $\mathbf{v_m} = v_r\hat{r}+v_\theta\hat{\theta}$. In all hydrodynamic cases, the balance is primarily between the Reynolds and viscous stresses, with the meridional circulation providing a positive radial flux near the boundaries of the CZ and negative at mid-depth, but aligning with the Reynolds stress latitudinally.

\subsection{MHD Convection and Mean Flow Structures}
\begin{figure}
	\centering
	\includegraphics[width=1.0\linewidth]{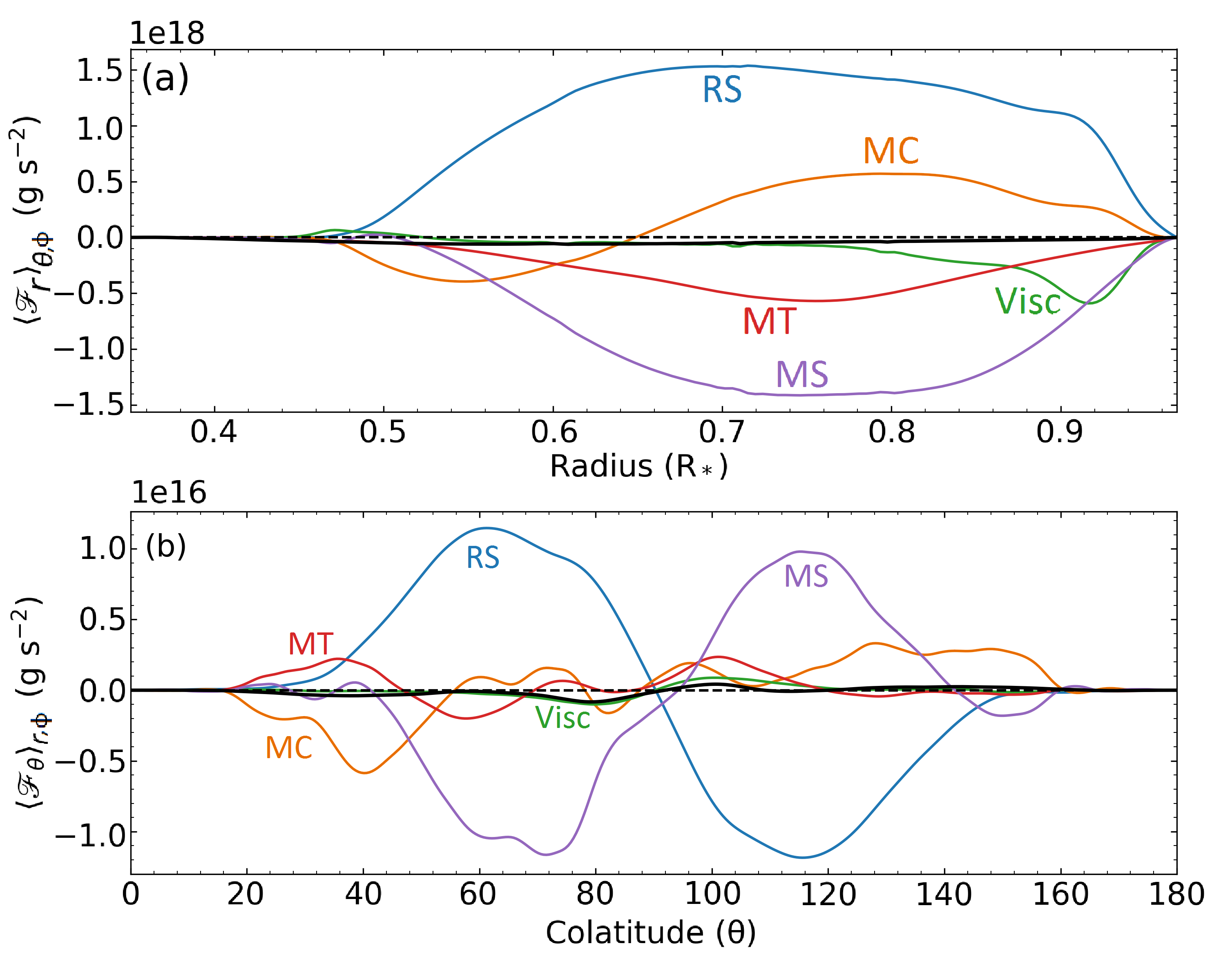}
	\caption{(a) Spherically- and time-averaged radial torque balance in the dynamo case D2t. The Maxwell stress replaces the viscous as the primary counterpart to the Reyndolds stress. (b) Cone- and time-averaged latitudinal torque balance in case D2t. The Reynolds stress is balanced by the Maxwell stress, and the meridional transport is asymmetric.}
	\label{fig:mhdamom}
\end{figure}

With the introduction of magnetism to the system through the admission of dynamo action, the flows in our models are self-consistently reshaped until they arrive at a new balance. Although the large-scale magnetism achieved in our models generally undergoes complicated time-evolution, we seek to produce a representative summary of their behavior through a combination of snapshots and long time-averages.

Representative snapshots of radial velocity in the near-surface convection achieved in cases D1t, D2, D2t, and D2ta are presented in Figure \ref{fig:mhdconv}(a), along with their patterns near mid-depth in Figure \ref{fig:mhdconv}(b). Although the character and amplitude of the convective flows in the MHD cases are very similar to those of their hydrodynamic counterparts near the equator, the isotropic cells found near the poles have severely diminished flow speeds in the presence of magnetism. In the linear theory of rotating, magnetized Rayleigh-Benard convection, sufficiently strong fields may increase the critical Rayleigh number $\mathrm{R_{a,c}}$ \citep{chandra61}. Studies of anelastic convective morphology in rotating spheres (e.g., \citealt{hindman18}) indicate that the rolling equatorial modes are the first to be destabilized at low supercriticality. This mirrors the trend among the hydrodynamic cases of shrinking polar convection with faster rotation, another factor which increases $\mathrm{R_{a,c}}$. We also note that the effect of a mid-latitude band of reduced convection seen in cases H1t, H2, H2t, and H4t (Figure \ref{fig:flows}a) is less pronounced in the magnetic cases, disappearing entirely in cases D1 and D1t, where the equatorial rolls blend seamlessly into the only slightly weakened polar cells.
\begin{figure*}[p]
	\centering
	\includegraphics[height=0.88\textheight]{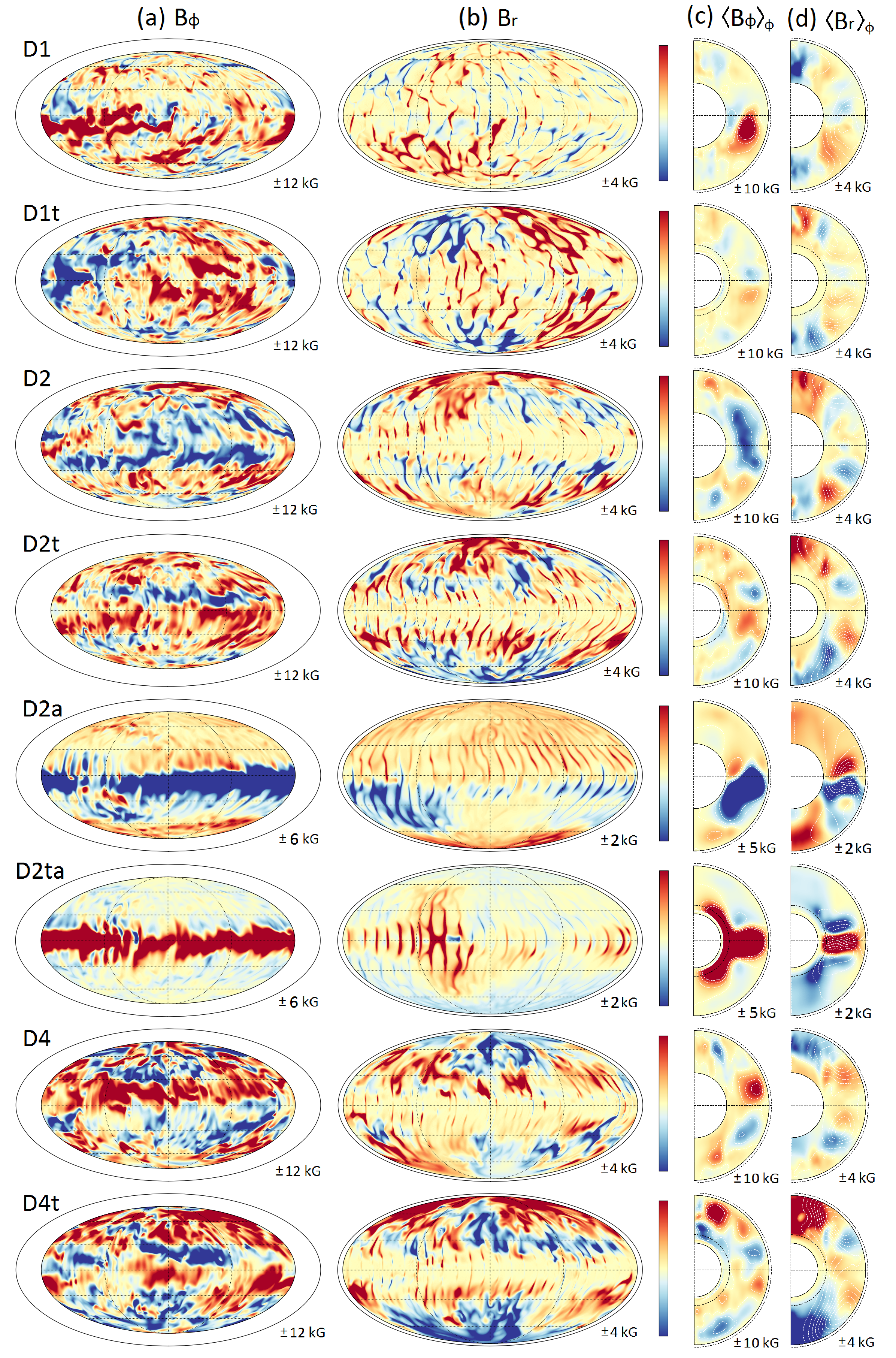}
	\caption{Representative snapshot samples of some of the magnetic structures achieved in each of our MHD models. Shown are (a) $B_\phi$ near mid-depth ($r=0.83R_*$) and (b) $B_r$ near the surface ($r=0.97R_*$) in Mollweide projection. Field amplitude and axisymmetry tend to increase with rotation rate. Near-surface fields tend to be significantly stronger for stars containing tachoclines, as quantified in Table \ref{tab:torque}. (c) $\langle B_\phi\rangle_{\phi,t}$ and (d) $\langle B_r \rangle_{\phi,t}$ are averaged in longitude and 2 years in time, over intervals indicated on Figure \ref{fig:B_evo}. Contours of the poloidal streamfunction are plotted over the radial field as white dashed lines.}
	\label{fig:magnetism}
\end{figure*}

The turbulent magnetic fields realized in all dynamo cases save D2a and D2ta lead to a dramatic reduction in the total differential rotation, presented for cases D1t, D2, D2t, and D2ta in Figure \ref{fig:mhdconv}(c). For the first two cases, and indeed all cases wih $\mathrm{P_{rm}}=4$, the star now rotates nearly as a solid body, with only a small band of prograde zonal flow remaining in a shallow layer at the equator. As a result, the rotational shear present in the tachocline is also severely reduced for these models. Considering the torque balance of D2t, shown in Figure \ref{fig:mhdamom}, it is clear that the Reynolds stress is now balanced primarily by the Maxwell stress, defined as 
\begin{equation}
\langle \mathbf{\mathcal{F}_\mathrm{MS}}\rangle_\phi = \frac{1}{4\pi}r\sin\theta\langle \mathbf{B_m}'B_\phi'\rangle_\phi \;,
\end{equation}
where $\mathbf{B_m} = B_r\hat{r}+B_\theta\hat{\theta}$. The viscous angular momentum transport has all but vanished, as one might expect for a sphere in near-solid-body rotation. The transport due to torques by the mean magnetic field, defined as
\begin{equation}
\langle \mathbf{\mathcal{F}_\mathrm{MT}}\rangle_\phi = \frac{1}{4\pi}r\sin\theta\langle \mathbf{B_m}\rangle_\phi \langle B_\phi\rangle_\phi \;,
\end{equation}
aligns with the Maxwell stress radially, but contributes little to the latitudinal torque balance. In cases D2a and D2ta, enhanced electrical resistivity limits the growth of small-scale magnetic fields, and thus the Maxwell stress is of insufficient amplitude to fully balance the Reynolds stress. This yields a rotational contrast with $\Delta\Omega$ roughly 50\% larger in D2ta than that of D2 and D2t, with the balance completed by the consequent viscous stresses. 

The meridional circulations, shown for D1t, D2, D2t, and D2ta in Figure \ref{fig:mhdconv}(d), have also been severely disrupted by the strong axisymmetric magnetic fields realized in our models. Excluding case D2ta which arrived at a time-steady solution, the complicated evolution of the magnetic fields leads to largely disorganized and asymmetric circulation cells which gradually merge and morph as the configuration of the global fields changes. In the steady case, D2ta, magnetic feedbacks reshape the circulations of its precursor H2t into two large cells outside the tangent cylinder in each hemisphere, with four narrow, axially-aligned 
cells within.

\section{Magnetic Fields}
We next turn to the magnetic fields built by dynamo action in our eight MHD simulations, the broad results of which are summarized in Table \ref{tab:magnetism}, analyzing the influence a tachocline has at various rotation rates within the regime of strong rotational constraint. Unless otherwise noted, time averages of the magnetic properties of these models are calculated over an interval of two years within a single half-cycle should the dynamo be cycling, or over the same interval as the hydrodynamic variables if the magnetism is steady.

\subsection{Magnetic Field Structures}
In Figure \ref{fig:magnetism}, we present snapshots of both $B_\phi$ and $B_r$, as well as meridional-plane views of azimuthal- and time-averages of the magnetic fields achieved in each of our magnetic models which are representative of the structures they tend to build over the course of the simulations. From just these narrow glimpses into the full complexity realized over the life of the simulations, it is clear that the magnetic structures possible within these models are highly diverse.  

Taking cases D2 and D2t as central to assessing the effects of a tachocline in our survey, clear similarities emerge between the two classes of models. At the times shown, the toroidal field structure in D2 is dominated by a single monolithic wreath spanning both hemispheres, whereas D2t has two wreaths of opposite polarity flanking the equator. These are only snapshots, however, and in general, the symmetry or asymmetry of the sense of the toroidal field across the equator does not seem to be correlated with the presence of a tachocline in our models. In the D2t, these two wreaths are not parallel to the equator, but rather tipped slightly poleward. They also tend to be asymmetric, rather than having roughly equivalent axisymmetric amplitudes in each hemisphere. In both cases, toroidal field strengths in the central wreaths peak in the vicinity of 20 kG, whereas their axisymmetric mean components reach about 10 kG for D2t and 12 kG for D2. Common between the two cases, we observe less structured fields at high latitude, with peak strengths of roughly 16 kG and means on the order of 4 kG. 

Qualitative differences between cases D2 and D2t begin to emerge, however, when considering the fields realized at the base of the CZ. Case D2t contains two additional wreaths found in the tachocline, directly below their mid-CZ counterparts. While the structures within the CZ are thoroughly modulated by the turbulent convection, those in the tachocline are far more laminar, and achieve both mean and peak strengths of roughly 8 kG. We find that the tachocline provides a reservoir for the toroidal field, as can be seen to varying degrees for all tachocline models in Figure \ref{fig:magnetism}. In solar models, the shear of the tachocline is thought to provide a mean field $\Omega$-effect for converting poloidal to toroidal field, and indeed, we find that the dominant inductive source term for these deep fields is that of the mean shear, which we discuss in detail in Section \ref{sec:anal}. 

An interesting behavior emerges in models D1 and D2a, which both produce extremely strong fields extending from the equator into only one hemisphere, while the other hemisphere remains magnetically barren by comparison. In the case of D2a (and D2ta, but in both hemispheres), the enhanced electrical resistivity can be used to explain the absence of turbulent fields away from the main wreaths, however no such argument can be applied to case D1. These hemispheric configurations could possibly be understood in terms of a quasi-linear interaction between a symmetric mode (e.g., what is observed in D2 at the selected time), and an antisymmetric one (e.g., in D1t) of roughly equivalent amplitude. Such models have been explored for the Sun as potential explanations for the Maunder minimum (e.g., \citealt{knobloch96}, \citealt{raynaud16}), yielding hemispheric modes similar to those we observe with the right choice of parameters. Due to the highly nonlinear nature of our solutions, however, it is difficult to know whether these combinations are realistic to expect. A marginally nonlinear analysis of the dynamo equations may provide a more concrete interpretation, but such an undertaking is beyond the scope of this work.


\subsection{Time-Dependence of Magnetic Structures}

\begin{figure*}
	\centering
	\includegraphics[width=0.9\linewidth]{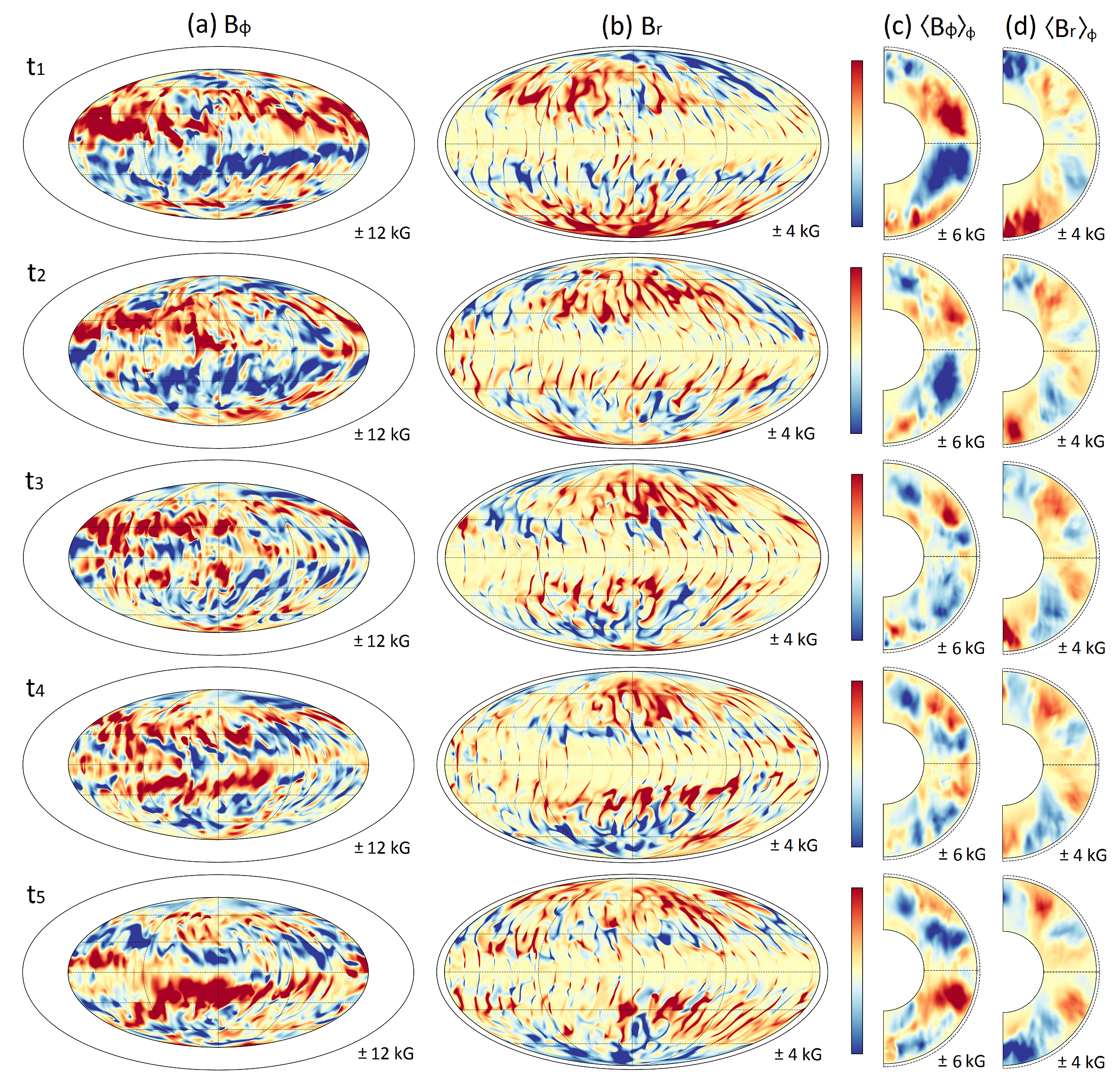}
	\caption{(a) Snapshots of $B_\phi$ near mid-depth ($r=0.77R_*$) and (b) $B_r$ near the surface ($r=0.96R_*$) in case D2 at five instants spanning 5.7 years as it undergoes a polarity reversal. The times selected fall within an interval marked on Figure \ref{fig:B_evo}. (c) Instantaneous azimuthal averages $\langle B_\phi \rangle_\phi$ and (d) $\langle B_r \rangle_\phi$ in the meridional plane.}
	\label{fig:reversal}
\end{figure*}

\begin{figure*}
	\centering
	\includegraphics[width=0.95\linewidth]{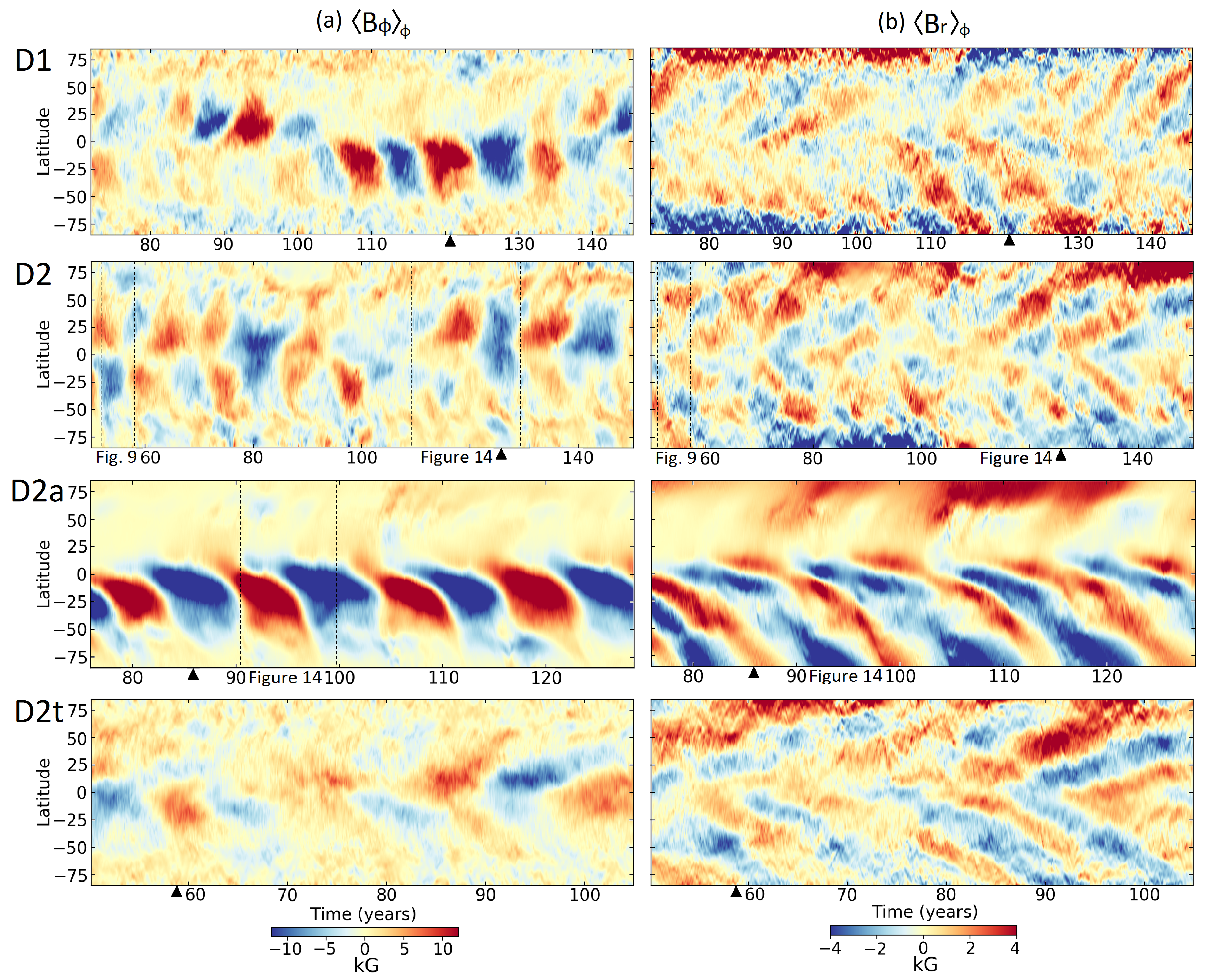}
	\caption{(a) Longitude-averaged azimuthal $\langle B_\phi\rangle_\phi$ and (b) radial magnetic fields $\langle B_r\rangle_\phi$ near mid-depth ($r=0.68R_*$), plotted against latitude and time for a selection of models. The toroidal fields tend to form a single strong structure on either side of the equator, reaching out to $\pm 50^\circ$, which may be symmetric, antisymmetric, or asymmetric, as the dynamo wanders between configurations. Radial fields tend to form chevron-like structures as they emerge near the equator and gradually migrate poleward through the life of a cycle. Intervals sampled in Figures \ref{fig:reversal} and \ref{fig:omega} are marked with dashed lines, and the central times of 2-year averaging intervals are shown with $\blacktriangle$.}
	\label{fig:B_evo}
\end{figure*} 

Although we have chosen to present representative snapshots of the magnetic fields achieved in each of our models in Figure \ref{fig:magnetism}, these structures are constantly shifting and evolving for all cases except D2ta. To explore the complicated time-evolution of the magnetic fields realized in these models, we first consider the morphologies of the longitudinally-averaged azimuthal $\langle B_\phi \rangle_\phi$ and radial $\langle B_r \rangle_\phi$ fields, which are plotted near mid-depth in time-latitude space for a selection of models in Figure \ref{fig:B_evo}. All models save D2ta and D4t exhibit regular polarity reversals in their magnetic fields in the bulk of the CZ. Representative Mollweide slices of $B_\phi$ and $B_r$ as well as their instantaneous longitude-averages $\langle B_\phi\rangle_\phi$ and $\langle B_r\rangle_\phi$ are shown in Figure \ref{fig:reversal} for one such reversal early in the evolution of case D2, here in an antisymmetric configuration. \textbf{Over the course of the reversal shown, beginning around time $t_2$, new axisymmetric fields emerge in a shallow band near the surface, between latitudes of roughly $\pm25^\circ$.} This location corresponds to the site of rotational shear remaining in the CZ from the strong differential rotation established in the hydrodynamic models. The emergence of these new fields displaces the existing wreaths toward their respective poles, where they may persist for several cycles before eventually breaking up. 

For each of the magnetic cases in Figure \ref{fig:B_evo}, the system migrates between symmetric, anti-symmetric, and asymmetric modes with respect to reflection across the equator. The asymmetric feature is particularly prominent in cases D1 and D2a, and seems to be migrating between hemispheres on a period roughly six times longer than the basic polarity reversals in case D1. In the framework of an interaction between quasi-linear modes, this beat frequency would imply that the periods of the symmetric and anti-symmetric modes are near the 6:7 resonance, though it would be difficult to identify which mode is faster without a careful analysis of their eigenfunctions. For D2a, the large-scale axisymmetric fields remain isolated to the southern hemisphere for the duration of the simulation. Quasi-linearly, this would suggest that the symmetric and antisymmetric modes have nearly identical periods for this model. 

While there is clear time dependence of the magnetic fields in case D4t (not shown), it is occuring over time scales which begin to become uneconomical to capture. Over the final 25 years of its run-time in which we consider it to be saturated, D4t undergoes no clear polarity reversals. Whether or not a cycle may emerge with further evolution is unclear from the runtime allowed. In case D4, we observe transitions between a wandering mode dominated by a single wreath with no clear period, and well-constrained cycling in an antisymmetric mode.

Considering the evolution of the radial fields in Figure \ref{fig:B_evo}(b), the cyclic behavior is clearly evident. Near mid-depth, we observe chevron-like structures which emerge near the equator and migrate poleward. The angular velocity of this feature's propagation seems to be nearly constant, though model dependent, near mid-depth and tends to be slow enough that roughly two chevrons will be nested at any given instant. This nesting leads to the predisposition we observe in the axisymmetric poloidal fields toward quadrupole and octupole modes, corresponding to symmetric and antisymmetric chevrons, respectively. Near the surface, the poleward propagation of the radial fields does not occur with fixed velocity. New axisymmetric fields first appear on either side of an exclusionary band reaching from the equator to approximately $\pm 10^\circ$ in latitude. Here their angular velocities are at first slower than the propagation deeper in the CZ, but break at latitudes of $\pm 25^\circ$ to become faster than the deep propagation and thus maintain the cycle time. Interestingly, the critical latitude for this behavior seems to correspond roughly to the latitudinal extent of the strong differential rotation remaining in the CZs of the dynamo models.

\begin{figure*}
	\centering
	\includegraphics[width=1.0\linewidth]{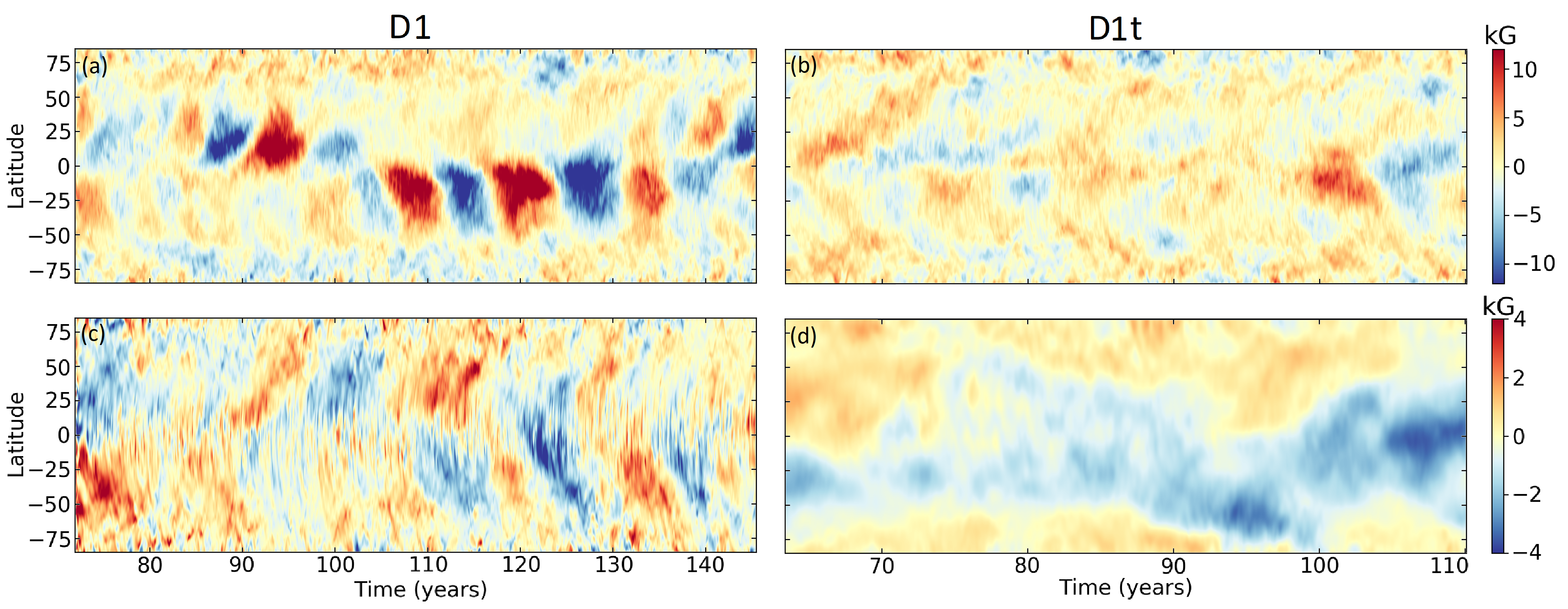}
	\caption{(a) Time-latitude diagrams of $\langle B_\phi \rangle_\phi$ near mid-depth ($r=0.68R_*$) in cases D1 and (b) D1t. (c) Time-latitude diagrams of $\langle B_\phi \rangle_\phi$ near the base of the CZ ($r=0.44R_*$) in cases D1 and (d) D1t.}
	\label{fig:bcz}
\end{figure*}
        
\begin{table*}[ht]
\centering
\begin{tabular*}{\textwidth}{l @{\extracolsep{\fill}} cccccc}
\hline \hline
Case & CKE & DRKE & PME & TME & FME & Period \\ \hline
D1 & 1.652 $\pm$ 0.223 & 2.618 $\pm$ 0.446 & 0.084 $\pm$ 0.020 & 0.383 $\pm$ 0.228 & 2.775 $\pm$ 0.319 & 11.0 $\pm$ 1.4 \\
D1t & 1.665 $\pm$ 0.241 & 1.135 $\pm$ 0.218 & 0.064 $\pm$ 0.019 & 0.121 $\pm$ 0.061 & 3.329 $\pm$ 0.386 & 8.8 $\pm$ 0.7 \\
D2 & 1.242 $\pm$ 0.164 & 1.276 $\pm$ 0.212 & 0.110 $\pm$ 0.038 & 0.361 $\pm$ 0.183 & 3.007 $\pm$ 0.330 & 12.3 $\pm$ 0.8 \\
D2t & 1.169 $\pm$ 0.302 & 1.009 $\pm$ 0.156 & 0.127 $\pm$ 0.049 & 0.304 $\pm$ 0.105 & 3.026 $\pm$ 0.300 & 14.0 $\pm$ 1.9\\
D4 & 0.874 $\pm$ 0.123 & 1.228 $\pm$ 0.175 & 0.209 $\pm$ 0.109 & 0.415 $\pm$ 0.159 & 4.143 $\pm$ 1.030 & 22.0 $\pm$ 4.2\\
D4t & 0.829 $\pm$ 0.112 & 1.143 $\pm$ 0.077 & 0.468 $\pm$ 0.100 & 0.417 $\pm$ 0.087 & 4.254 $\pm$ 0.564 & - \\
D2a & 1.488 $\pm$ 0.205 & 2.953 $\pm$ 0.722 & 0.141 $\pm$ 0.042 & 1.298 $\pm$ 0.442 & 1.169 $\pm$ 0.175 & 12.4 $\pm$ 2.4 \\
D2ta & 1.433 $\pm$ 0.174 & 2.612 $\pm$ 0.127 & 0.156 $\pm$ 0.044 & 1.944 $\pm$ 0.217 & 1.174 $\pm$ 0.129 & - \\ \hline \hline
\end{tabular*}
    \caption{Properties of the dynamos realized in each of our MHD models. Energy densities are averaged over full simulation duration after maturity is reached, and are given in units of $10^6$ erg cm$^{-3}$. Definitions for each energy density are presented in Section \ref{sec:anal}. Uncertainties represent one standard deviation. Cycle periods are reported in units of years where available, with uncertainties determined by the FWHM of the periodogram.}
    \label{tab:magnetism}
\end{table*}

\begin{figure}
	\centering
	\includegraphics[width=1.0\linewidth]{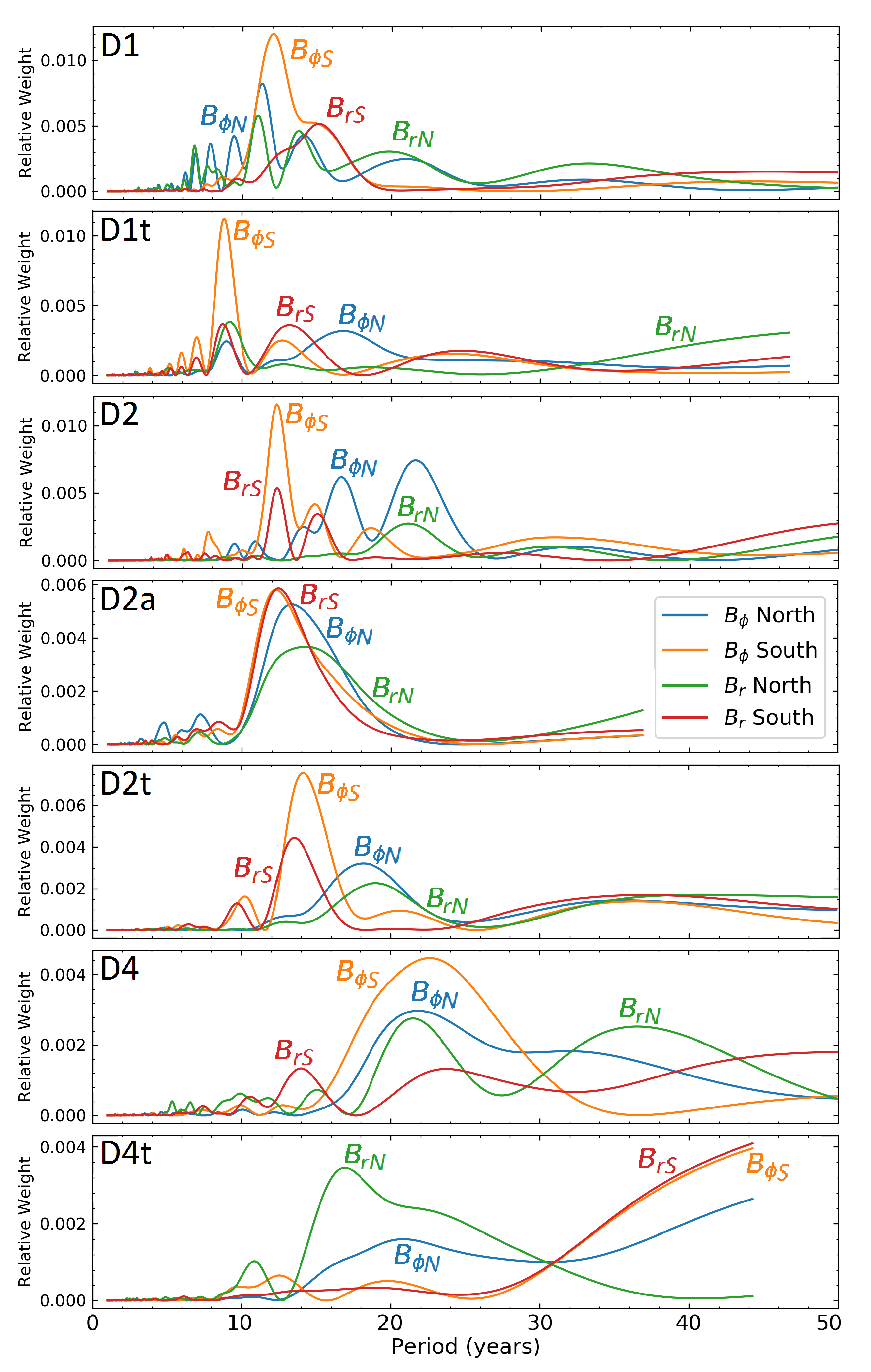}
	\caption{Lomb-Scargle periodograms of the radial and azimuthal magnetic fields at depth $0.68R_*$ averaged over latitudinal intervals from $10^\circ$ to $40^\circ$. \textbf{Due to the fine time-resolution of our data, all peaks visible in the spectra represent statistically significant fluctuations.} Spectral power at long time-scales, as seen prominently in the fields of D4t, represent a non-detection of a cycle within the captured simulation time. }
	\label{fig:periods}
\end{figure} 
 
At all rotation rates, we find that the fields at the bottom of the CZ are significantly affected by the presence of a tachocline. We present in Figure \ref{fig:bcz} time-latitude diagrams of $\langle B_\phi \rangle_\phi$ at the base of the CZ and near mid-depth for cases D1 and D1t. From this, one may see that the deep fields of D1t are much more coherent in time. At times the tachocline fields of case D1t correlate well with the mid-CZ fields, but it is equally likely to find them in the opposite sense. By way of contrast, the deep fields of case D1 tend to reflect all the reversals occurring above, though they may be out of synchronization by up to a quarter of a cycle. This trend holds for all of our paired models: cases D1t, D2t, and D4t all have significantly more coherent deep fields which tend to maintain their state for much longer times than those in their purely convective counterparts.

To analyze cycle lengths, we compute Lomb-Scargle periodograms of the axisymmetric toroidal and radial fields of each hemisphere near mid-depth and averaged in latitude from $10^\circ$ to $40^\circ$, shown in Figure \ref{fig:periods}. The periods we identify are reported in the final column of Table \ref{tab:magnetism}. A full cycle consists of two polarity reversals, therefore a failed reversal in one or both hemispheres can lead to the periodogram emphasizing a harmonic of the true period. For instance, a single failed reversal increases the duration of the cycle by 50\%, while doubling the duration of a single phase, and thus leads to power in both the 3:2 and 2:1 harmonics. This can be observed for case D2, whose broad peaks at ~17 and ~22 years are rough harmonics of the ~12 year period. Interestingly, D2a possesses the same cycle period as case D2. This suggests that the two models may share a reversal mechanism, despite the differences in resistivity and overall character between them. 

From the results of this analysis, we can identify two major effects on the temporal evolution of fields in these simulations caused by the presence of a tachocline. Firstly, the tachocline acts as a stabilizer for the global mean fields, helping to reduce the signature of high frequency noise. Secondly, coupling between the tachocline fields and the mid-CZ dynamo can lead in some cases to cycle periods which are much longer (perhaps even indefinite) than those that could be achieved without the tachocline. 

\subsection{Time-Steady Magnetic Structures}
While case D2a and all the models running with $\mathrm{P_{rm}}=4$ except case D4t yielded cycling solutions of various timescales, case D2ta persisted in a steady state for most of its simulated life. Depicted in Figure \ref{fig:magnetism}, the fields in D2ta assumed a quadrupolar form featuring a single wreath at the equator in the CZ melding into a monolithic sheet of magnetic field of the same sense forming in the tachocline and spreading upward through the base of the CZ. Through the core of the wreath and tachocline sheet, we observe $\langle B_\phi \rangle_\phi$ in the vicinity of 30 kG, with peaks exceeding 50 kG in the wreath.

This configuration of field persists with no significant evolution for more than 30 years, until buoyant magnetic loops rising from the tachocline in the southern hemisphere upset the symmetry of the system and cause it to begin cycling with a period of approximately 13 years in an asymmetric manner very similar to what we find for case D2a. We reserve further discussion of that transition for a future paper.

\subsection{Analyzing Inductive Processes}
\label{sec:anal}
\begin{figure*}
	\centering
	\includegraphics[width=1.0\linewidth]{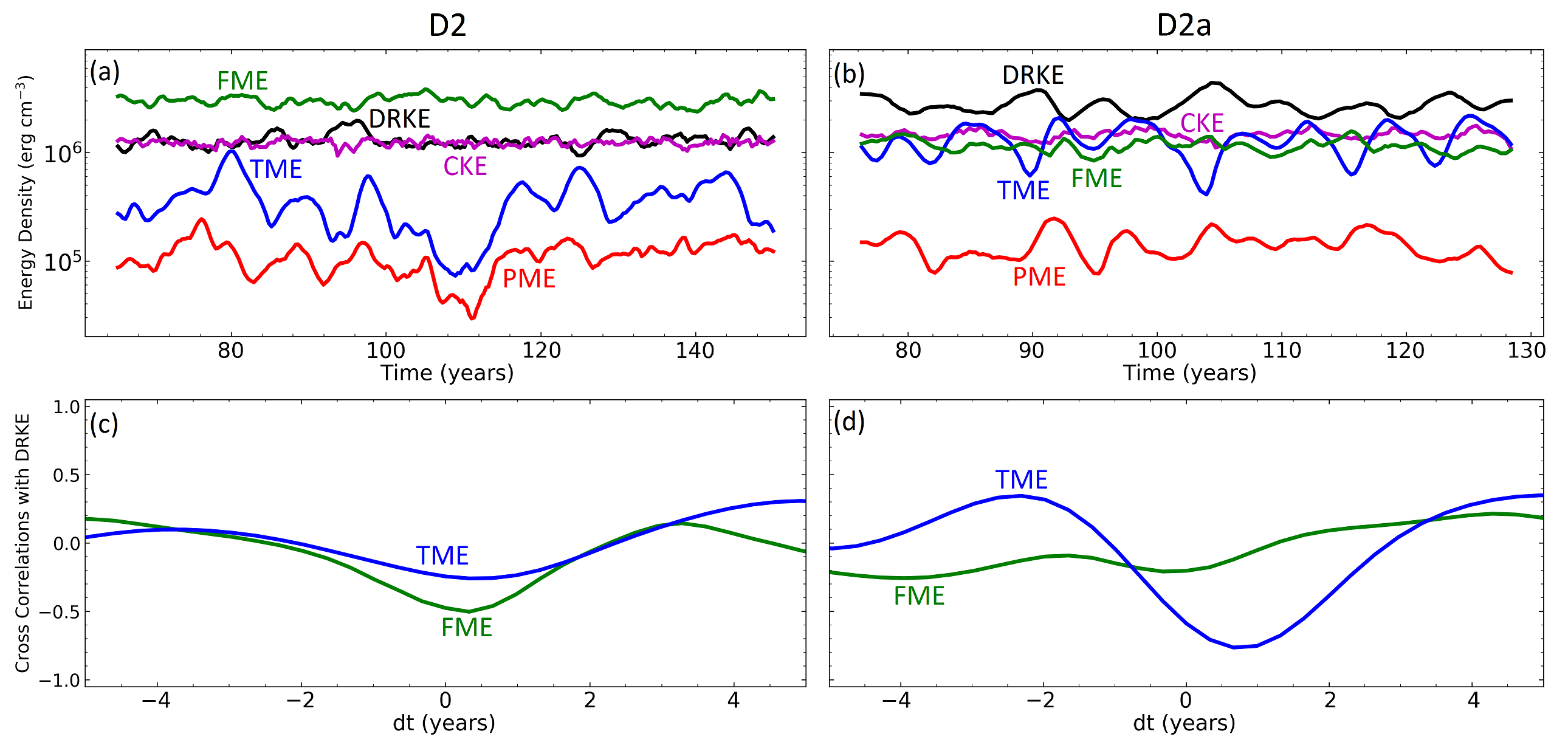}
	\caption{(a) Time evolution of globally averaged energy densities DRKE (black), CKE (green), TME (blue), PME (red), and FME (green) in case D2. (b) Time evolution of globally averaged energy densities in case D2a. DRKE and TME are much larger than in D2, while FME is reduced. (c) Cross-correlations of TME (blue) and FME (green) with DRKE in model D2. Positive $dt$ indicates that fluctuations of the labeled field precede those of DRKE. (d) Cross-correlations of TME (blue) and FME (green) with DRKE in model D2a. }
	\label{fig:energy}
\end{figure*}

We next examine the evolution of global averages of the energies associated with dynamo action in cases D2 and D2a, which together capture qualitatively all the inductive behaviors we observe in the cycling cases. Time-traces of these energies are presented in Figure \ref{fig:energy}, along with their normalized cross-correlations. These are the convective kinetic energy (CKE), differential rotation kinetic energy (DRKE), poloidal magnetic energy (PME), toroidal magnetic energy (TME), and fluctuating magnetic energy (FME), defined as 

\begin{equation}
\mathrm{CKE}=\frac{1}{2}\bar{\rho}(\mathbf{v}-\langle\mathbf{v}\rangle_\phi)^2\;,
\end{equation}
\begin{equation}
\mathrm{DRKE}=\frac{1}{2}\bar{\rho}\langle v_\phi \rangle_\phi^2\;,
\end{equation}
\begin{equation}
\mathrm{PME}=\frac{1}{8\pi}(\langle B_r \rangle_\phi^2+\langle B_\theta \rangle_\phi^2)\;,
\end{equation}
\begin{equation}
\mathrm{TME}=\frac{1}{8\pi}\langle B_\phi \rangle_\phi^2\;,
\end{equation}
\begin{equation}
\mathrm{FME}=\frac{1}{8\pi}(\mathbf{B}-\langle\mathbf{B}\rangle_\phi)^2\;.
\end{equation}

While its large amplitude means that the CKE is undoubtedly a significant player in the overall dynamics of the system, we observe no consistent correlations between its flucuations and those of the DRKE or magnetic energies in our models, and so we will exclude it from further discussion. In case D2, and indeed the rest of the $\mathrm{P_{rm}}=4$ models, there is a clear anticorrelation between the $\langle$DRKE$\rangle$ and $\langle$FME$\rangle$ which is broken only during particularly strong peaks of the $\langle$PME$\rangle$ and TME$\rangle$. The $\langle$FME$\rangle$ tends to attain its local extrema fewer than 100 days before the $\langle$DRKE$\rangle$, implying that the two enjoy a parasitic relationship. As the non-axisymmetric fields draw energy from the differential rotation, their Maxwell stresses grow stronger, consequently quenching their power source in the DRKE. With the enhanced electrical resistivity in cases D2a and D2ta, the amplitude of the $\langle$FME$\rangle$ is considerably reduced. This allows these models to maintain a much stronger differential rotation, and consequently drive more energy in the axisymmetric PME and TME than their less resistive counterparts. In case D2a, fluctuations of the $\langle$DRKE$\rangle$ are anti-correlated instead with changes in the $\langle$TME$\rangle$ through a mean magnetic torque. 

The stark difference in the energy distributions between these two sets of models suggests that their fundamental dynamo processes may differ as well. To begin exploring these processes, we first break up the induction term from Equation \ref{eqn:induction} into contributions from shear, advection, and compression, which take the forms $(B\cdot\nabla)v$, $(v\cdot\nabla)B$, and $B(\nabla\cdot v)$, respectively. Taking advantage of the anelastic condition and steady background, compressive induction can be recast in the form $-Bv_r\frac{dln\hat{\rho}}{dr}$. For tractability, such analyses are typically only presented as they apply to the axisymmetric components of the magnetic field, retaining only ``mean" $\langle x\rangle_\phi \langle y\rangle_\phi$ and ``fluctuating" $\langle  x'y' \rangle_\phi$ components. Details of this derivation can be seen in Appendix A of \citet{brownsteady}. For our purposes, we further dot these source terms with a unit vector aligned with the local mean field. This yields equations whose sign denotes enhancement or reduction of the mean field, rather than the sense of field being generated. Under this formalism, the dominant source terms affecting the mean magnetic fields in our models are

\begin{figure*}
	\centering
	\includegraphics[width=1.0\linewidth]{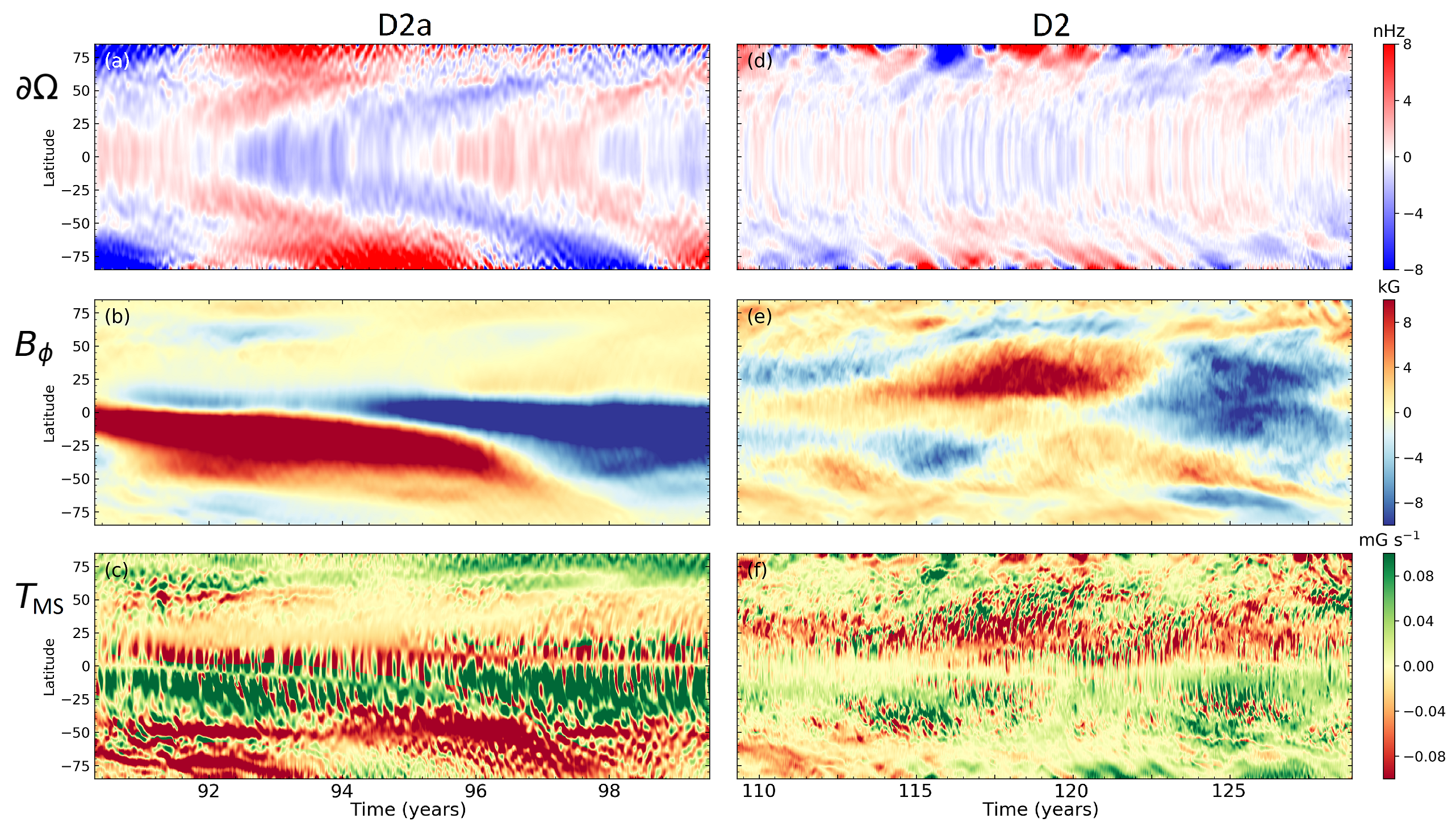}
	\caption{(a,d) Time-latitude diagrams near mid-depth of fluctuations in the local rotation rate $\partial\Omega$ relative to an average in time and longitude for cases D2a and D2, respectively. Red denotes faster rotation, and blue slower. (b,e) Time-latitude diagrams near mid-depth of $\langle B_\phi \rangle_\phi$ in cases D2a and D2, respectively. (c,f) Time-latitude diagrams near mid-depth of $T_{MS}$ in cases D2a and D2, respectively. Green denotes enhancement of the local $\langle B_\phi \rangle_\phi$, whereas red tones indicate destruction of field.}
	\label{fig:omega}
\end{figure*}

\begin{equation}
S_\mathrm{MS} = \langle \hat{B} \rangle_\phi \cdot [(\langle B \rangle_\phi \cdot\nabla)\langle v \rangle_\phi]\;,
\label{eqn:ind1}
\end{equation}
\begin{equation}
S_\mathrm{MA} = \langle \hat{B} \rangle_\phi \cdot [(\langle v \rangle_\phi \cdot\nabla)\langle B \rangle_\phi]\;,
\end{equation}
\begin{equation}
S_\mathrm{FS} = \langle \hat{B} \rangle_\phi \cdot \langle(B' \cdot\nabla)v' \rangle_\phi\;,
\end{equation}
\begin{equation}
S_\mathrm{FA} = \langle \hat{B} \rangle_\phi \cdot \langle(v' \cdot\nabla)B' \rangle_\phi\;,
\end{equation}
\begin{equation}
S_\mathrm{FC} = -\langle \hat{B} \rangle_\phi \cdot \langle B'v_r'\frac{dln\rho}{dr} \rangle_\phi\;,
\end{equation}
\begin{equation}
S_\mathrm{RD} = -\langle \hat{B} \rangle_\phi \cdot \nabla\times\eta\nabla\times\langle B \rangle_\phi\;.
\label{eqn:ind2}
\end{equation}

We further break down these sources into their contributions to the toroidal and poloidal field components (e.g. $S_\mathrm{MS} = T_\mathrm{MS} + P_\mathrm{MS}$). We present in Figure \ref{fig:omega} time-latitude diagrams of differential rotation fluctuations $\delta\Omega = \langle \Omega \rangle_\phi - \langle \Omega \rangle_{\phi,t}$ and $T_{MS}$ over one cycle of cases D2a and D2. From panel (a), it is clear that D2a is undergoing strong and well-defined torsional oscillations, which set its cycle period. Throughout the cycle shown, the rotational shear serves to consistently reinforce the magnetic wreath within its core, and break apart fields that drift to higher or lower latitudes. Of particular interest is the observation that while the large-scale magnetic fields involved in the cycle of D2a are generally restricted to the southern hemisphere, their feedback on the differential rotation is symmetric across the equator. This we attribute to a rhythmic disruption of the meridional circulations in the northern hemisphere by the formation of fields near the equator. Because these circulations transport angular momentum, they are able to carry the signature of the torsional oscillation even where there are no strong fields present.  

\begin{figure*}
	\centering
	\includegraphics[width=1.0\linewidth]{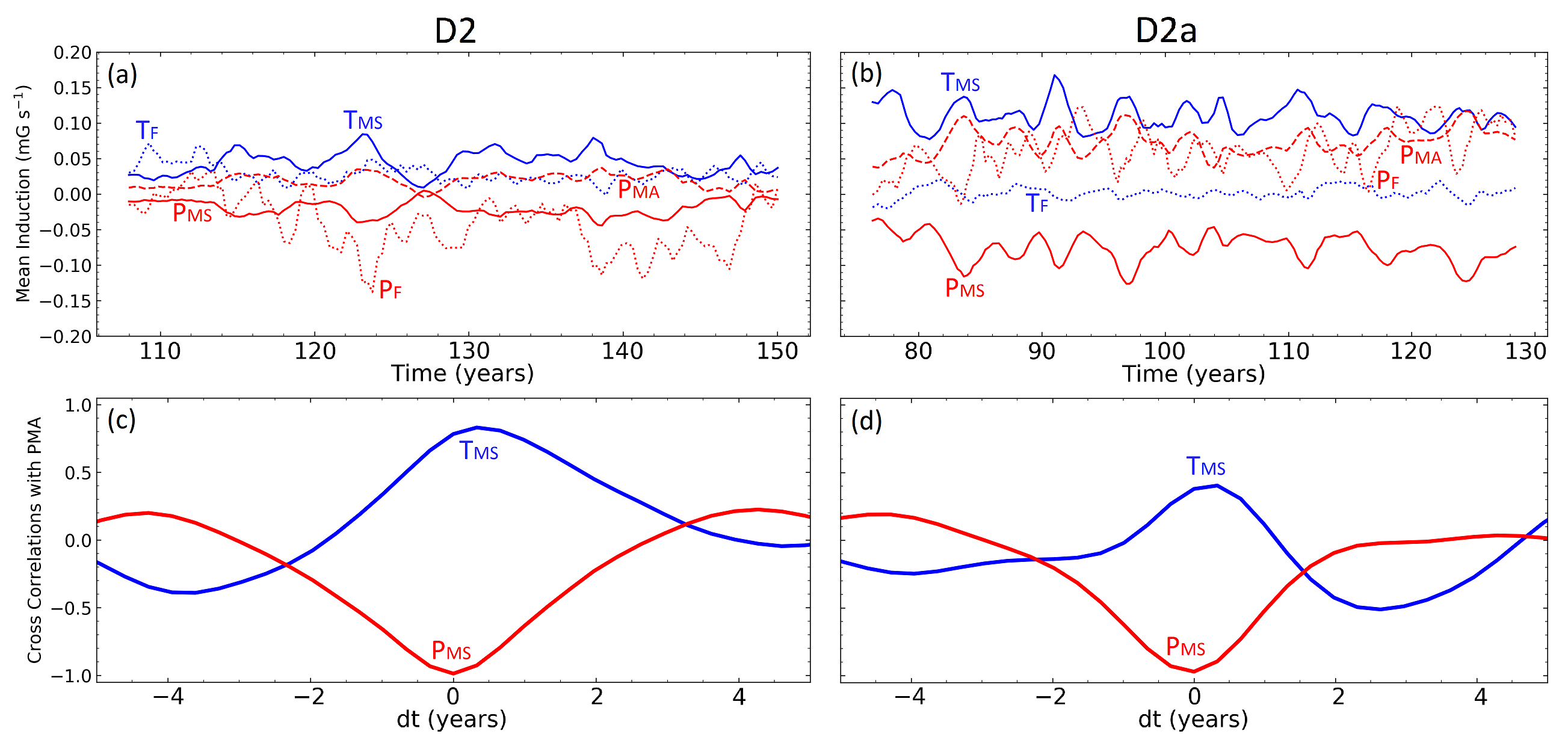}
	\caption{(a) Time evolution of globally-averaged induction by mean shear (solid), mean advection (dashed), and combined non-axisymmetric processes $S_\mathrm{F} = S_\mathrm{FS}+S_\mathrm{FA}+S_\mathrm{FC}$ (dotted) for the toroidal (blue) and poloidal (red) fields over a shortened interval of D2. (b) Time evolution of globally-averaged inductive source terms in case D2a. While mean field terms obey similar trends to D2, fluctuating induction is strikingly different in character. (c) Cross-correlations of $T_\mathrm{MS}$ (blue) and $P_\mathrm{MS}$ (red) with $P_\mathrm{MA}$ in model D2. (d) Cross-correlations of $T_\mathrm{MS}$ (blue) and $P_\mathrm{MS}$ (red) with $P_\mathrm{MA}$ in model D2a.}
	\label{fig:alpha}
\end{figure*}

While these oscillations are much weaker and less coherent in case D2 (Figure \ref{fig:omega}d,f), they are not wholly absent. Here the oscillations are evident in the rotation rate at latitudes lower than approximately 35$^\circ$, but may be disrupted as they propagate poleward. Furthermore, the $T_{MS}$ of D2 does not uniformly support the existing wreaths. At the time shown, $T_{MS}$ acts to damp the mean fields of the northern hemisphere while enhancing those in the south, and does not form a confining sheath as is the case in D2a. Despite these differences, the shared mean reversal time of cases D2 and D2a suggests that torsional oscillations may still be responsible for setting this timescale even in the more turbulent models. 

To further explore differences in the dynamo processes operating within our models, we consider global averages of the inductive source terms dominating magnetic evolution in cases D2 and D2a, presented in Figure \ref{fig:alpha}. We note that in both cases, $T_{MS}$ is correlated with $P_{MA}$, which is itself nearly equal and opposite to $P_{MS}$, whereas $T_{MA}$ is negligibly small. These effects together complete the picture for the more orderly dynamo process based around mean-shear. Advective transport gathers poloidal field in places where it can then be transformed through shear into toroidal field. 

However, the induction of the mean fields through non-axisymmetric processes is extremely different for cases D2 and D2a. In the more resistive model D2a, $T_F$ does not contribute significantly to the overall dynamics, and $P_F$ has a positive amplitude comparable to $P_{MA}$ and $P_{MS}$. In case D2, however, $T_{F}$ is comparable to $P_{MA}$, and $P_{F}$ is extremely negative. While the chaotic nature of these fluctuating inductions make their exact impacts difficult to quantify, their significance to the magnetic field evolution in the set of models with $\mathrm{P_{rm}=4}$ reinforces the interpretation that the strong, turbulent fields we achieve can obfuscate the underlying process responsible for the cycle period in the torsional oscillations.

Within the tachoclines of all of our models containing them, magnetic fields are somewhat removed from the turbulent fluid motions of the CZ, and they attain much more laminar configurations similar to those observed for the more resistive cases D2a and D2ta. The time-evolution of these fields is then dominated by the large-scale $S_{MS}$ and $S_{MA}$ terms, and thus we might expect that their long-term behavior may also be well described in terms of torsional oscillations. However, although the $\partial\Omega$ signal of the mid-CZ reversals is able to penetrate into the upper reaches of the tachocline, it does not in most cases lead to a reversal of the fields contained there. Within the CZ, magnetic structures left unsupported by a waning $S_{MS}$ are broken apart and recycled by turbulent convection and resistive decay, allowing structures of the opposite sense the space and energy to form. The timescale for this resistive decay can be estimated by $\tau_\eta=L_B^2/\eta$, with $L_B^2$ the average size of magnetic structures, defined from the magnetic field spectrum as:

\begin{equation}
L_B = \pi r \frac{\sum_\ell B_\ell}{\sum_\ell \ell B_\ell}
\end{equation}

At a depth of $r=0.8R_*$, we have a decay time comparable to the reversal time, at $\tau_\eta=$ 7.1 years. Within the RZ of our models, however, the timescale for this process is greater than 500 years, and so the structures there are typically able to survive long enough to be refreshed in the waxing phase of the oscillation. We note that this timescale is enhanced in the RZ in part due to our reduced $\eta$ there, but the larger $L_B$ there would preserve this separation of scales even for a uniform $\eta$.
 
\subsection{Emergent Fields and Spin-Down}

\begin{figure*}
	\centering
	\includegraphics[width=0.95\linewidth]{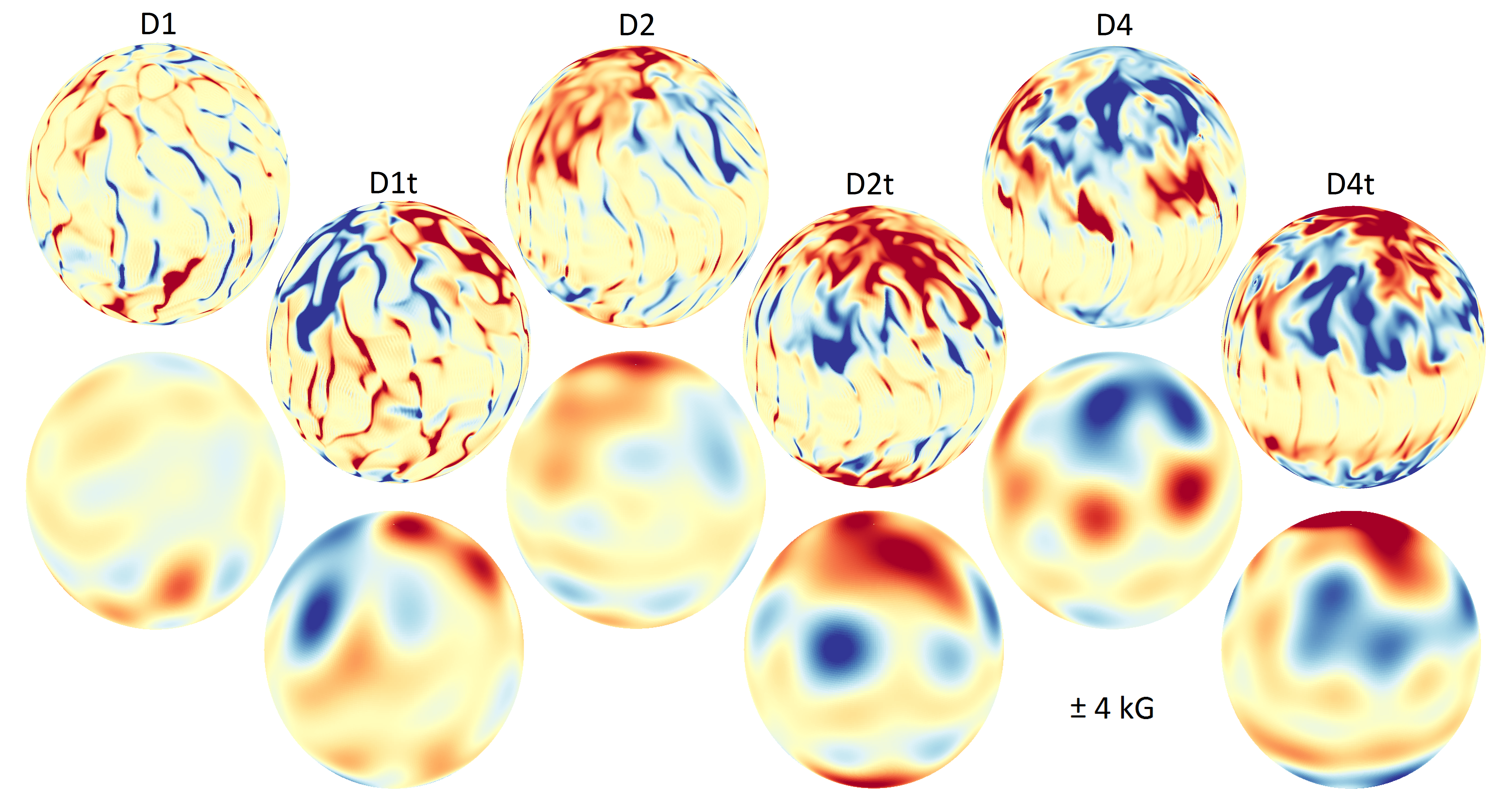}
	\caption{Near-surface $B_r$ at full spatial resolution (above) and under a low-pass spatial filter (below) for our dynamo models with $\mathrm{P_{rm}}=4$. When filtered, only magnetic field components with spherical harmonic order $l\leq12$ are shown. The mean unsigned radial field of the fully resolved slices are reported in Table \ref{tab:torque}.}
	\label{fig:emergent}
\end{figure*}

The poloidal fields realized within the various simulations reveal a clear difference between models that contain tachoclines and those that do not in their coupling between field strengths and rotation rate. Representative near-surface $B_r$ and averaged poloidal fields are shown for each model in Figure \ref{fig:magnetism}(b) and (d), respectively. From Table \ref{tab:torque}, one can see that the near-surface radial fields are both stronger and fill a larger fraction of the sphere in models which contain a tachocline. This is especially evident when filtering the surface fields to only components with spherical harmonic order $l\leq12$, presented in Figure \ref{fig:emergent}. This treatment has been shown by \citet{yadav15} to approximate reasonably well the fields which may be inferred through Zeeman Doppler Imaging (ZDI) observations of real stars.

Considering the axisymmetric poloidal fields of the pure-CZ models in Figure \ref{fig:magnetism}(d), we can see that the radial fields of cases D1, D2, and D4 are all quite similar. All are largest in the polar region where they reach peaks on the order of 5 kG near mid-depth. When we consider the models containing tachoclines, however, they show strikingly different mean field strengths and configurations at different rotation rates. In D1t, the mean radial fields are of comparable strength to D1, but tend to peak nearer to the surface. Case D2t possess stronger field strengths than D2, which are configured with more order. D4t has by far the strongest and most widespread mean radial fields of any model, with polar caps reaching down to $45^\circ$ and mean strengths there approaching 10 kG. D4 and D4t also show significant low latitude radial fields of comparable strength that reach the surface of the domain, which are largely absent in the slower-rotating models.

If we seek to explain the the magnetized spin-down of these early M-stars, then we must consider not just the fields present within the star, but those that may extend outward into the surrounding environment. Thus, we turn our attention specifically to the radial fields present at the top of the domain for each model, as in Figure \ref{fig:emergent}. We must keep in mind that the top of our domain is about $0.03R_*$ from the true surface, and so the structures present will face a certain amount of disruption before potentially breaching the photosphere. Certain morphologies are common among all the simulations. At low latitudes, the radial fields are concentrated in the downflow lanes. There, they occasionally build up enough strength to locally suppress the convection in a way analogous to star spots. At higher latitudes, the convection is far less organized and the radial magnetism stronger, and so we see a much higher filling factor for strong fields.

\begin{table}
\centering
\begin{tabular*}{\linewidth}{l @{\extracolsep{\fill}} ccc}
\\ \hline \hline
Case & $B_r^{surf}$ (G) & n & $\dot{J}$  \\ \hline
D1 & 757 $\pm$ 19 & 5.68 $\pm$ 0.07 & 1.364 $\pm$ 0.143 \\
D1t & 1099 $\pm$ 26 & 5.37 $\pm$ 0.08 & 2.954 $\pm$ 0.321 \\
D2 & 977 $\pm$ 18 & 5.43 $\pm$ 0.06 & 2.370 $\pm$ 0.226 \\
D2t & 1194 $\pm$ 20 & 5.30 $\pm$ 0.05 & 3.072 $\pm$ 0.190 \\
D4 & 1007 $\pm$ 40 & 5.04 $\pm$ 0.11 & 5.218 $\pm$ 0.885 \\
D4t & 1462 $\pm$ 30 & 4.49 $\pm$ 0.06 & 11.78 $\pm$ 0.934 \\ \hline \hline
\end{tabular*}
    \caption{RMS surface field, flux weighted complexity, and average magnetic torque affecting each pair of models in the core sequence. Time-averaging is done over the full mature lifetime of each model, with the uncertainty representing one standard deviation of the mean. For each pair, the torque is significantly larger in the model containing a tachocline.}
    \label{tab:torque}
\end{table}

Much work has been done by stellar wind modeling groups to explore and parameterize the torques associated with increasingly complex magnetic field geometries. In particular, we turn here to \citet{garraffo}, which reports a scaling for the angular momentum loss rate of

\begin{equation}
\dot{J}(n,B) = \dot{J}_{dip}(B)[4.05e^{-1.4n}+ \frac{n-1}{60Bn}]\;,
\end{equation}
where $n$ is a flux-weighted average of the spherical harmonic order of the surface magnetic field:

\begin{equation}
n = \frac{\sum_llF_l}{\sum_{l}F_l}\;.
\end{equation}

Separately, the torque exerted by a purely dipolar field has been shown to scale approximately with the surface field strength $\dot{J}_{dip}\sim B$ (e.g. \citealt{winds}). Taking these together, we present in Table \ref{tab:torque} non-dimensional estimates for the relative torques affecting each of our pairs of models. To maintain contact with the field geometries in \citet{garraffo}, we restrict our focus in this analysis to only field components with $l\leq12$, which should be the dominant contributors to spin-down regardless. In each pair of models considered, we find larger $B$ and smaller $n$ in the model containing a tachocline, which combine for significantly larger angular momentum loss rates in these stars. These results are consistent, at least qualitatively, with the assertion that the presence of a tachocline in early M-stars, or conversely its absence in late M-stars, may be a contributing factor to the distinct transition observed in magnetic lifetime across the tachocline divide.

\section{Conclusions}
We have compared the results of global MHD simulations of 13 different models of fast-rotating M2-like stars. Five of these models included only the CZ within their computational domain, while the other eight extended deeper into the star to incorporate portions of the underlying RZ and tachocline. In doing so, we have arrived at a number of conclusions surrounding the flows of these stars, the magnetic fields they produce through dynamo action, and how each of these may be influenced by the presence of a tachocline.

While our hydrodynamic models were able to establish strong, solar-like differential rotations, these shearing flows are strongly quenched in the presence of magnetism. We find that the degree of quenching is greater for our less resistive models, which poses an interesting question for how this effect may extend to the parameter spaces of real stars: is the degree of $\Omega$-quenching dependent on the scale of the smallest magnetic structures relative to the overall system ($\mathrm{R_{em}}$), or relative to the smallest fluid structures ($\mathrm{P_{rm}}$)? More work is currently underway to determine an answer, which may yield predictions for the differential rotation observable on real M-dwarfs.    

We have shown that the CZs of fast-rotating early M-dwarfs are perfectly capable of generating and organizing strong toroidal fields with or without an underlying tachocline of shear, though the presence of a tachocline may encourage stronger fields for faster rotators. Even in cases with very little discernable differential rotation, we have found that the primary mechanism for the reversal of the global fields in these models is the torsional oscillation, as seen in Figure \ref{fig:omega}. 

As has been found in simulations of the Sun, we have shown that the tachocline can provide a reservoir for the magnetic fields produced in the bulk of the CZ, as well as acting as a site for a secondary dynamo. Coupling between this reservoir and the turbulent mid-CZ dynamo can regulate the reversals of the global field, or in some extreme cases prevent them altogether.

We have found that the tachocline helps to enhance and organize near-surface poloidal fields onto larger spatial scales for fast-rotating early M-dwarfs. According to stellar wind models, this may create a favorable condition for these stars to shed angular momentum more rapidly through their magnetized winds.

While few experiments of this type have been conducted for early M-dwarfs, the more extensively studied parameter spaces in the solar regime have proven to house a rich diversity of behavior. More work is currently underway to examine the local sensitivities of our models in parameter space, and thus to assess the robustness of these conclusions concerning the features of deep convective shells with underlying tachoclines.

Furthermore, we have compared the properties of only M2-like models, and have not in this work bridged the tachocline divide. To truly argue that tachoclines allow early M-dwarfs to spin down more readily than late, we must compare these results to those of similar experiments for FC stars. While a small body of FC simulations exist, and in the case of \citet{yadav15} produce dipole-dominant surface fields of greater magnitude than were found in any of our simulations, they all exist in disparate enough regions of parameter space from our own study that direct comparisons of such quantities are not fruitful. Instead, we must extend the modeling choices and parameterizations of this work to new models of FC stars to enable a proper comparison. Additionally, this study has only considered how the role of a tachocline varies with the rotation rate and to a much lesser degree the magnetic Prandtl number. Further experiments exploring higher Rayleigh numbers or alternative representations of the sub-gridscale diffusion may shed more light on the robustness of our results, or perhaps reveal interesting new behaviors.

\software
{MESA \citep{mesasw},
Rayleigh \citep{rayleighsw}}

\acknowledgements
{We thank Kyle Augustson, Benjamin Brown, Sacha Brun, and Brad Hindman, as well as the reviewer for helpful advice in developing this work. We thank Nick Featherstone for authoring the Rayleigh code and for his assistance with its use, as well as the Computational Infrastructure for Geodynamics which is funded by the National Science Foundation. Computational resources were provided by the NASA High-End Computing (HEC) program through the Pleiades supercomputer at NAS in the Ames Research Center. This work was primarily supported by NASA Astrophysical Theory Program grant NNX17AG22G, and also partly by Heliophysics grants 80NSSC18K1127 and NNX16AC92G.}

\end{document}